\documentclass[a4paper,11pt]{article}
\pdfoutput=1
\usepackage[utf8]{inputenc}
\usepackage[english]{babel}
\usepackage{notation}

\usepackage[normalem]{ulem}
\definecolor{myGreen}{RGB}{54, 150, 45}
\usepackage{placeins}
\usepackage{tikz-cd}
\usetikzlibrary{math}
\usepackage{physics}

\begin{document}

\begin{titlepage}
\begin{flushright}
UWThPh 2024-27
\end{flushright}

\vskip 2cm
\begin{center}
{\Large{\textbf{Classification of Minimal Abelian Coulomb Branches}}}

\vspace{15mm}
{{\large 
Antoine Bourget${}^a$, Quentin Lamouret${}^a$

Sinan Moura Soys\"uren${}^b$, 
Marcus Sperling${}^b$}} 
\\[5mm]
\noindent {${}^a$\em  Institut de physique théorique, Université Paris-Saclay, CEA, CNRS, \\ 91191, Gif-sur-Yvette, France}\\
Email: {\tt antoine.bourget@ipht.fr}, \\ {\tt quentin.lamouret@ipht.fr}
\\[5mm]
\noindent {${}^b$\em Fakultät für Physik, Universität Wien,\\
Boltzmanngasse 5, 1090 Wien, Austria,}\\
Email: {\tt sinan.moura.soysueren@univie.ac.at},\\ \tt{marcus.sperling@univie.ac.at}
\\[15mm]
\end{center}

\begin{abstract}

Obtaining the classification of 3d $\Ncal=4$ quivers whose Coulomb branches have an isolated singularity is an essential step in understanding moduli spaces of vacua of supersymmetric field theories with 8 supercharges in any dimension. 
In this work, we derive a full classification for such Abelian quivers with arbitrary charges, and identify all possible Coulomb branch geometries as quotients of $\mathbb{H}^n$ by $\urm(1)$ or a finite cyclic group. 

We give two proofs, one which uses the \emph{decay and fission} algorithm, and another one relying only on explicit computations involving 3d mirror symmetry. 
In the process, we put forward a method for computing the 3d mirror of any $\urm(1)^r$ gauge theory, which is sensitive to discrete gauge factors in the mirror theory. This constitutes a confirmation for the decay and fission algorithm.

\end{abstract}

\end{titlepage}

\tableofcontents

\section{Introduction}

Supersymmetric quantum field theories in spacetime dimension $d=3,4,5,6$ with 8 supercharges display a remarkable diversity, and can be approached from a variety of points of view, from weak coupling Lagrangian descriptions to inherently non-perturbative definitions via geometry or brane setups in string theory. However, there is a sense in which all this diversity can be reduced to a list of elementary building blocks. This uses the fact that such theories generically admit large continuous spaces $\mathcal{M}$ of (supersymmetric) vacua. The action of the R-symmetry algebra (which contains a summand $\mathfrak{su}(2)_H$ for any such $d$) on $\mathcal{M}$ allows to identify a distinguished branch, the Higgs branch $\mathcal{M}_H$, which is a singular hyper-K\"ahler space --- a symplectic singularity.

One characteristic feature of a symplectic singularity is the existence of a stratification into a partially ordered finite set of symplectic leaves \cite{Kaledin:2006}. This allows for the graphical representation into a Hasse diagram. 
Physically, this corresponds to the various ways a theory can be Higgsed, or said differently, the various renormalization group (RG) flows triggered by a specific class of relevant deformations. Particularly relevant are the the so-called minimal degenerations, which correspond to elementary Higgsings, or equivalently Higgs/Coulomb branch RG-flows. By definition, every Higgs branch can be broken down into the minimal degenerations which appear in the Hasse diagram.  These are the building blocks alluded to above. 

The question is then, is it possible to classify all transverse slices that appear as minimal degenerations? 
For the nilpotent orbits of semi-simple Lie algebras, the classification is known \cite{kraft1980minimal,Kraft1982,FU20171} and encompasses (essentially) only Kleinian surface singularities and minimal nilpotent orbit closures. However, for more general symplectic singularities the situation is not yet resolved. In fact, some previously unknown isolated symplectic singularities have been found recently in \cite{bellamy2023new,Bourget:2021siw,Bourget:2022tmw}.

Historically, the Higgs branches of \emph{Lagrangian} theories with 8 supercharges are a physical manifestation of the hyper-K\"ahler quotient \cite{Hitchin:1986ea} and in some cases have been understood in the context of Nakajima quiver varieties \cite{Nakajima:1994nid}. But the description of Higgs branches of non-Lagrangian theories, like all 5d and 6d superconformal field theories (SCFTs), or Argyres-Douglas theories in 4d, requires more general tools. 

One such tool is the Coulomb branch of 3d $\Ncal=4$ SCFTs, which under mild assumptions is also a symplectic singularity. Famously, 3d $\Ncal=4$ mirror symmetry \cite{Intriligator:1996ex} is an infrared duality that exchanges Higgs and Coulomb branch of a mirror dual pair of 3d $\Ncal=4$ SCFTs. Physically, this is exciting, because Coulomb branches are severely affected by quantum corrections in contrast to 3d Higgs branches, which are entirely governed by classical physics. To remedy this situation, a variety of physics approaches has been developed: Abelianisation \cite{Bullimore:2015lsa}, Coulomb branch quantisation \cite{Dedushenko:2018icp}, and Hilbert series \cite{Cremonesi:2013lqa}.
Mathematically, this unveiled a new class of holomorphic symplectic singularities --- the 3d $\Ncal=4$ Coulomb branch --- which, motivated by \cite{Cremonesi:2013lqa}, led Braverman, Nakajima, and Finkelberg to pursue a rigorous mathematical definition \cite{Braverman:2016wma}.

In the magnetic quiver program  (see \cite{Cremonesi:2015lsa,Ferlito:2017xdq,Cabrera:2018jxt,Cabrera:2019izd,Bourget:2019rtl,Cabrera:2019dob} and subsequent works), one uses string theory to realize the Higgs branch of the theory under study as the Coulomb branch of a 3d $\Ncal=4$ quiver, or generalization thereof involving so-called non-simply laced edges \cite{Cremonesi:2014xha}. When a magnetic quiver is found for the Higgs branch of a theory, one can obtain the Hasse diagram using known algorithms, such as quiver subtraction \cite{Bourget:2019aer,Bourget:2020mez,Bourget:2022ehw,Bourget:2022tmw}, or decay and fission \cite{Bourget:2023dkj,Bourget:2024mgn}. Here, we turn this around and use decay and fission to identify magnetic quivers for the elementary building blocks. More precisely, our aim is to push towards a classification of \emph{isolated conical symplectic singularities} (ICSSs) in the class of symplectic singularities realized as 3d $\Ncal=4$ Coulomb branches of unitary quiver theories. An ICSS is realized by a unitary quiver whose Coulomb branch has exactly two leaves, the trivial and the non-trivial one --- such Coulomb branches are denoted as \emph{minimal Coulomb branches} here, and the corresponding quivers are called \emph{stable quivers}. In \cite{Bourget:2023dkj,Bourget:2024mgn}, the list of stable quivers with at most 3 nodes (and bifundamental or non-simply laced edges) has been obtained. We now aim at the general case. In the present article, we completely solve this problem for quivers that contain only $\urm(1)$ gauge nodes (and arbitrary charges). The extension to arbitrary $\urm(n)$ nodes will be performed in a subsequent work. We note that, as a rule of thumb based on previous partial explorations, we expect there to be fewer and fewer stable quivers as the gauge groups become big. Hence the Abelian case, while simple conceptually, is the richest in terms of number of minimal Coulomb branches.

\paragraph{Notation.} 
The following notation is introduced, for $n,k$ positive integers and $\sigma \in \mathbb{Z}^n$: 
\begin{align}
  h_{n,k,\sigma} &:= (\mathbb{H}^n/\Z_k)^{[\sigma]}
  \label{eq:h-def}
  \\
  \overline{h}_{n-1,\sigma} &:= (\mathbb{H}^n/\urm(1))^{[\sigma]}
  \label{eq:hbar-def}
\end{align}
i.e.\ a discrete $\Z_k$ and continuous $\urm(1)$ hyper-K\"ahler quotient of the space $\mathbb{H}^n$ carrying the charge vector $\sigma$ under the $\Z_k$ and $\urm(1)$ action, respectively. These generalize the $h_{n,k} \equiv  h_{n,k,(1,\ldots,1)}$ and $ \overline{h}_{n,1} \equiv  \overline{h}_{n,(1,\ldots,1)}  =a_{n}$ spaces of \cite{Bourget:2021siw}. The same notation $h_{n,k,\sigma}$ for orbifolds of $\mathbb{H}^n$ was appreciated in \cite{Grimminger:2024doq}.

\subsection{Summary of results and plan of the paper} 

We now summarize the main results of the papers, which constitute on the one hand a full classification of stable quivers with $\urm(1)$ gauge nodes, and on the other hand a full classification of the geometries that can be obtained from these stable quivers. 
We first give the list of stable quivers with $\urm(1)$ gauge nodes: 

\noindent\fbox{
\parbox{\textwidth}{\vspace*{-10pt}
\paragraph{Proposition 1. } 
A unitary 3d $\mathcal{N}=4$ Abelian quiver gauge theory with \emph{well-defined length} has a Coulomb branch which is an $n$-dimensional ICSS if and only if it is one of
\begin{enumerate}
    \item[$(i)$] A chain of $n+1$ vertices 
        \begin{equation}
        \raisebox{-.5\height}{\begin{tikzpicture}
    \tikzset{node distance=2.5cm};
    \node (1a) [gauge,label={[align=center]below:\small{$1$}}] {};
    \node (1b) [gauge,right of=1a,label={[align=center]below:\small{$1$}}] {};
    \node (1dot) [right of=1b] {$\cdots$};
    \node (1e) [gauge,right of=1dot,label={[align=center]below:\small{$1$}}]{};
    \node (1f) [gauge,right of=1e,label={[align=center]below:\small{$1$}}]{};
    \draw[line width=2pt,gray, decoration={markings, mark=at position 0.2 with {\arrow[scale=0.7]{>}}, mark=at position 0.9 with {\arrow[scale=0.7]{<}}}, postaction={decorate}] (1a)to node[midway, above, black] {\small{$(\ell_{1},k_{1})$}} (1b);
    \draw[line width=2pt,gray, decoration={markings, mark=at position 0.2 with {\arrow[scale=0.7]{>}}, mark=at position 0.9 with {\arrow[scale=0.7]{<}}}, postaction={decorate}] (1b)to node[midway, above, black] {\small{$(\ell_{2},k_{2})$}} (1dot);
    \draw[line width=2pt,gray, decoration={markings, mark=at position 0.2 with {\arrow[scale=0.7]{>}}, mark=at position 0.9 with {\arrow[scale=0.7]{<}}}, postaction={decorate}] (1dot) to node[midway, above, black] {\small{$(\ell_{n-1},k_{n-1})$}} (1e);
    \draw[line width=2pt,gray, decoration={markings, mark=at position 0.2 with {\arrow[scale=0.7]{>}}, mark=at position 0.9 with {\arrow[scale=0.7]{<}}}, postaction={decorate}] (1e)to node[midway, above, black] {\small{$(\ell_{n},k_{n})$}} (1f);
\end{tikzpicture} }   \label{eq:chainGeneral}
    \end{equation}
    with $\mathrm{gcd}(\ell_1 , k_n) > 1$ and $\mathrm{gcd}(\ell_i , k_j) = 1$ for all other $1 \leq i \leq j \leq n$.  
    \item[$(ii)$]  A cycle of $n+1$ vertices  
    \begin{equation}    \raisebox{-.5\height}{    \begin{tikzpicture}
    \tikzset{node distance=2cm};
    \node (1a) [gauge,label={[align=center]above:\small{$1$}}] {};
    \node (1b) [gauge,right of=1a,label={[align=center]above:\small{$1$}}] {};
    \node (1dot) [right of=1b] {$\cdots$};
    \node (1e) [gauge,right of=1dot,label={[align=center]above:\small{$1$}}]{};
    \node (1f) [gauge,right of=1e,label={[align=center]above:\small{$1$}}]{};
    \node (B1a) [gauge,above of=1dot,label={[align=center]below:\small{$1$}}] {};
    \draw[line width=2pt,gray, decoration={markings, mark=at position 0.5 with {\arrow[scale=0.7]{>}}, mark=at position 0.9 with {\arrow[scale=0.7]{<}}}, postaction={decorate}] (1a)to node[midway, below, black] {\small{$(\ell_{1},k_{1})$}} (1b);
    \draw[line width=2pt,gray, decoration={markings, mark=at position 0.2 with {\arrow[scale=0.7]{>}}, mark=at position 0.9 with {\arrow[scale=0.7]{<}}}, postaction={decorate}] (1b)to node[midway, above, black] {} (1dot);
    \draw[line width=2pt,gray, decoration={markings, mark=at position 0.2 with {\arrow[scale=0.7]{>}}, mark=at position 0.9 with {\arrow[scale=0.7]{<}}}, postaction={decorate}] (1dot)to node[midway, above, black] {} (1e);
    \draw[line width=2pt,gray, decoration={markings, mark=at position 0.2 with {\arrow[scale=0.7]{>}}, mark=at position 0.6 with {\arrow[scale=0.7]{<}}}, postaction={decorate}] (1e)to node[midway, below, black] {\small{$(\ell_{n-1},k_{n-1})$}} (1f);
    \draw[line width=2pt,gray, decoration={markings, mark=at position 0.4 with {\arrow[scale=0.7]{>}}, mark=at position 0.9 with {\arrow[scale=0.7]{<}}}, postaction={decorate}] (1a)to node[midway, above left, black,rotate=25] {\small{$(k_{n+1},\ell_{n+1})$\hspace*{-1cm}}} (B1a);
    \draw[line width=2pt,gray, decoration={markings, mark=at position 0.4 with {\arrow[scale=0.7]{>}}, mark=at position 0.9 with {\arrow[scale=0.7]{<}}}, postaction={decorate}] (1f)to node[midway, above right, black,rotate=-25] {\small{\hspace*{-1cm}$(k_{n},\ell_{n})$}} (B1a);
\end{tikzpicture} } \label{eq:quiverCircular}
    \end{equation}
with $\prod_i k_i = \prod_j \ell_j$ (``well-defined length'') and $\mathrm{gcd}(\ell_i , k_j) = 1$ for all pairs $(i,j)\in\Z_{n+1}^{2}$ such that $i-j \not\equiv 1,2$ mod $n+1$.  
\end{enumerate}}}

We provide two proofs of this statement in Sections \ref{Sec:Abelian_Chain} and \ref{Sec:Abelian_Cycle}. One makes use of the decay and fission algorithm, whilst the other relies solely on 3d mirror symmetry, as reviewed in Section \ref{Sec:Mirror_Construction}. The fact that the two proofs agree can also be seen as a confirmation of the validity and usefulness of the decay and fission algorithm.

Note that two different quivers can encode the same geometry. We now give the ICSS that are realized by the above quivers, along with a ``canonical'' representative: 

\noindent\fbox{
\parbox{\textwidth}{\vspace*{-10pt}
\paragraph{Proposition 2. } The $n$-dimensional ICSS that are Coulomb branches of 3d $\Ncal=4$ gauge theories with gauge group $\urm(1)^n$ are exactly
\begin{enumerate}
    \item[$(i)$] $h_{n,k,\sigma}$ for $k \in \mathbb{Z}_{\geq 2}$ and $\sigma = (\sigma_1 , \dots , \sigma_n) \in (\mathbb{Z}_k^\ast)^n$, which can be realized as\footnotemark
    \begin{equation}
        h_{n,\delta,\sigma} = \mathcal{M}_C \left[ \raisebox{-.5\height}{ \begin{tikzpicture}
    \tikzset{node distance=2.5cm};
    \node (1a) [flavour,label={[align=center]below:\small{$1$}}] {};
    \node (1b) [gauge,right of=1a,label={[align=center]below:\small{$1$}}] {};
    \node (1dot) [right of=1b] {$\cdots$};
    \node (1e) [gauge,right of=1dot,label={[align=center]below:\small{$1$}}]{};
    \node (1f) [gauge,right of=1e,label={[align=center]below:\small{$1$}}]{};
    \draw (1a)to node[midway, above, black] {} (1b);
    \draw[line width=2pt,gray, decoration={markings, mark=at position 0.2 with {\arrow[scale=0.7]{>}}}, postaction={decorate}] (1b)to node[midway, above, black] {\small{\hspace*{-.5cm}$\sigma_n / \sigma_{n-1}$}} (1dot);
    \draw[line width=2pt,gray, decoration={markings, mark=at position 0.2 with {\arrow[scale=0.7]{>}}}, postaction={decorate}] (1dot) to node[midway, above, black] {\small{\hspace*{-.5cm}$\sigma_3 / \sigma_2$}} (1e);
    \draw[line width=2pt,gray, decoration={markings, mark=at position 0.2 with {\arrow[scale=0.7]{>}}, mark=at position 0.9 with {\arrow[scale=0.7]{<}}}, postaction={decorate}] (1e)to node[midway, above, black] {\small{$(\sigma_2 / \sigma_1 , \delta)$}} (1f);
\end{tikzpicture}}  \right] \,  . \label{eq:chainReduced}
    \end{equation}
    Moreover, the Coulomb branch of any quiver of the form \eqref{eq:chainGeneral} can be rewritten in the form \eqref{eq:chainReduced}. The translation is $\delta=\mathrm{gcd}(\ell_1,k_{n})$ and $\sigma_i /  \sigma_{i-1}  = - \ell_i / k_{i-1}$ mod $\delta$ for $i=2,\dots,n$. 
    \item[$(ii)$]  $\overline{h}_{n,\sigma}$ for $\sigma = (\sigma_1 , \dots , \sigma_{n+1}) \in \mathbb{Z}^{n+1}$ such that for all $i,j$, $\mathrm{gcd}(\sigma_i , \sigma_j)=1$ which can be realized as 
    \begin{equation}
        \overline{h}_{n,\sigma} = \mathcal{M}_C \left[ \raisebox{-.5\height}{    \begin{tikzpicture}
    \tikzset{node distance=2cm};
    \node (1a) [gauge,label={[align=center]above:\small{$1$}}] {};
    \node (1b) [gauge,right of=1a,label={[align=center]above:\small{$1$}}] {};
    \node (1dot) [right of=1b] {$\cdots$};
    \node (1e) [gauge,right of=1dot,label={[align=center]above:\small{$1$}}]{};
    \node (1f) [gauge,right of=1e,label={[align=center]above:\small{$1$}}]{};
    \node (B1a) [gauge,above of=1dot,label={[align=center]below:\small{$1$}}] {};
    \draw[line width=2pt,gray, decoration={markings, mark=at position 0.5 with {\arrow[scale=0.7]{>}}, mark=at position 0.9 with {\arrow[scale=0.7]{<}}}, postaction={decorate}] (1a)to node[midway, below, black] {\small{$(\sigma_{1},\sigma_{3})$}} (1b);
    \draw[line width=2pt,gray, decoration={markings, mark=at position 0.2 with {\arrow[scale=0.7]{>}}, mark=at position 0.9 with {\arrow[scale=0.7]{<}}}, postaction={decorate}] (1b)to node[midway, above, black] {} (1dot);
    \draw[line width=2pt,gray, decoration={markings, mark=at position 0.2 with {\arrow[scale=0.7]{>}}, mark=at position 0.9 with {\arrow[scale=0.7]{<}}}, postaction={decorate}] (1dot)to node[midway, above, black] {} (1e);
    \draw[line width=2pt,gray, decoration={markings, mark=at position 0.2 with {\arrow[scale=0.7]{>}}, mark=at position 0.6 with {\arrow[scale=0.7]{<}}}, postaction={decorate}] (1e)to node[midway, below, black] {\small{$(\sigma_{n-1},\sigma_{n+1})$}} (1f);
    \draw[line width=2pt,gray, decoration={markings, mark=at position 0.4 with {\arrow[scale=0.7]{>}}, mark=at position 0.9 with {\arrow[scale=0.7]{<}}}, postaction={decorate}] (1a)to node[midway, above left, black,rotate=25] {\small{$(\sigma_{2},\sigma_{n+1})$\hspace*{-1cm}}} (B1a);
    \draw[line width=2pt,gray, decoration={markings, mark=at position 0.4 with {\arrow[scale=0.7]{>}}, mark=at position 0.9 with {\arrow[scale=0.7]{<}}}, postaction={decorate}] (1f)to node[midway, above right, black,rotate=-25] {\small{\hspace*{-1cm}$(\sigma_{1},\sigma_{n})$}} (B1a);
\end{tikzpicture} } \right] \, . \label{eq:circularReduced}
    \end{equation}
    Moreover, the Coulomb branch of \eqref{eq:quiverCircular} under the assumptions in Proposition 1 can be rewritten in the form \eqref{eq:circularReduced} with $\sigma_i = \gcd (\ell_{i+2} , k_i)$.  
\end{enumerate} }}

\footnotetext{Here division by $\sigma_i$ should be understood as multiplication by the inverse of $\sigma_i$ in the multiplicative group $\mathbb{Z}_k^\ast$. }

We have used a framing in \eqref{eq:chainReduced}. The relationship between framed and unframed quivers is elaborated upon in Section~\ref{Sec:Quiver_reduced_form}. Finally, the relevance of the assumption of well-defined length is discussed in Section~\ref{Sec:Abelian_Cycle_Extension}. 

\section{Unitary (magnetic) quivers FAQs}
\subsection{Lightning review of (magnetic) quivers and moduli spaces}
\label{Sec:setup}

Consider a 3d $\Ncal=4$ non-trivial, interacting superconformal field theory which, due to the superconformal algebra, enjoys an $\surm(2)_{\Higgs} \times \surm(2)_{\Coulomb}$ R-symmetry.  

\paragraph{Quiver theories.} Concretely, consider a 3d $\Ncal=4$ Lagrangian quantum field theory in the ultraviolet (UV) that flows to a non-trivial, interacting superconformal field theory (SCFT) in the infrared (IR). Among all such Lagrangian 3d $\Ncal=4$ theories, there exists a class of theories specified by an unoriented \emph{quiver} graph, which is composed of vertices and edges. That is, a vertex labeled by a (compact) Lie group $G$ denotes a 3d $\Ncal=4$ $G$ vector multiplet. An edge between two vertices $i$ and $j$ denotes a 3d $\Ncal=4$ hypermultiplet that transforms in the bifundamental representation of $G_i \times G_j$. This class is called quiver gauge theories. Among this class, there exists the subclass of \emph{unitary quiver gauge theories}, for which $G_i = \urm(n_i)$ for all $i$. This class is likely to be the most well-studied of all 3d $\Ncal=4$ SCFTs. \emph{Abelian unitary quiver theories} are then the subclass of unitary quivers for which all $G_i = \urm(1)$.

\paragraph{Higgs/Coulomb branch.} Any 3d $\Ncal=4$ SCFT comes with an $\surm(2)_{\Higgs} \times \surm(2)_{\Coulomb}$ R-symmetry, which then gives rise to two (maximal) hyper-K\"ahler branches --- called Higgs branch $\mathcal{M}_{\Higgs}$ and Coulomb branch $\mathcal{M}_{\Coulomb}$.  Important is also the notion of \emph{good} 3d $\Ncal=4$ quiver theories, which are such that there are no operators that violate the unitarity bound during the RG-flow into the IR. 
Henceforth, the considered class of theories is that of good, unitary, Abelian quiver gauge theories. (See further discussion below.)

As both maximal branches are singular, they admit \emph{symplectic singularities} \cite{Beauville:2000}. Such a structure is, in fact, accompanied by a finite stratification into symplectic leaves \cite{Kaledin:2006}.  These leaves are ordered by inclusion (of their closures). The transverse slice of a symplectic leaf, in the closure of a larger leaf, is itself a symplectic singularity. Thus, for two ordered leaves such that no other leaf is in between -- a so-called minimal degeneration -- the transverse slice $S$ is a minimal symplectic singularity $X_{\min}$. The latter means that the stratification of $S$ has only two leaves: the trivial (aka the origin) and the non-trivial leaf (whose closure is the entire space). Consequently, the singularity of such $X_{\min}$ is \emph{isolated} -- just at a single point, given by the origin. As these Higgs/Coulomb branches originate from SCFTs, the branches come equipped with a $\C^\ast$-action; in other words, they are \emph{conical}. 

Here, the focus is placed on identifying Abelian unitary quivers such that their Coulomb branches are \emph{isolated conical symplectic singularities} (ICSSs). As the focus is placed on the Coulomb branch of the quiver, the notion of a \emph{magnetic quiver} is used (see \cite{Cremonesi:2015lsa,Ferlito:2017xdq,Cabrera:2018jxt,Cabrera:2019izd,Bourget:2019rtl,Cabrera:2019dob} and subsequent works).

\paragraph{Class of Abelian quivers considered in this paper.} The focus of this paper is on the class of good 3d $\Ncal=4$ unitary Abelian quiver gauge theories with $n$-vertices and (generalized) edges corresponding to hypermultiplets with arbitrary integer-valued charges $(\ell,-k)$. In the language of unitary quiver theories, such general charges have also been called \emph{non-simply laced} edges \cite{Hanany:2001iy,Cremonesi:2014xha} (for either $k$ or $\ell$ equals $1$) or $(p,q)$-edges\footnote{In the convention of this paper, the charges $(l,-k)$ of a hypermultiplet would translate to a $(p,q)=(l,k)$ edge in that reference.} \cite{Grimminger:2024doq}.

As is customary for Abelian theories, one simply keeps track of these charges in the form of a \emph{charge matrix} $\rho$. In the present work, the convention is that its columns correspond to factors of the gauge group and its rows to hypermultiplets. More explicitly, it is the map induced from the representation morphism $G=\urm(1)^{n_G}\to \urm(1)^{n_H}$ on the co-character lattices:
\begin{align}
    \rho: \Z^{n_G} \to \Z^{n_H} \; ,
    \label{eq:charge_matrix_def}
\end{align}
where $G$ is the gauge group, $n_G$ the number of Abelian vector multiplets, and $n_H$ the number of hypermultiplets.
If the kernel $K = \operatorname{Ker}( G\to \urm(1)^{n_H})$ of the representation morphism contains a $\urm(1)$ subgroup, the associated gauge field decouples from the theory. The vacuum moduli space is then the product of the one for a free vector multiplet and the one of the rest of the theory. To avoid this decoupling of free degrees of freedom, one redefines the gauge group of the quiver theory to be the quotient $G = \urm(1)^{n_G}/K_0$ where $K_0$ is the connected part of the kernel $K$. This group $G$ can still have a finite subgroup that acts trivially. As is demonstrated below, taking this discrete subgroup into account is important to understand the structure of the vacuum moduli space. 
 
\paragraph{The notions of \emph{good} and \emph{balance}.}
\label{sec:good-balance}
A central question for 3d $\Ncal=4$ UV Lagrangian theories concerns their RG-flows into the IR. Does the original Lagrangian flow to an interacting CFT at the fixed point or do certain degrees of freedom decouple during this process?  
In superconformal theories, there exists a relation between the scaling dimension and the R-symmetry charges. Moreover, there exists a unitary bound (as a condition on the scaling dimension of operators) that determines whether a theory flows to an interacting SCFT or whether some degrees of freedom need to be removed. For 3d $\Ncal=4$ theories, these considerations led to a condition on the conformal dimension\footnote{In contrast to original reference \cite{Gaiotto:2008ak} and also \cite{Cremonesi:2013lqa}, the conformal dimensions is scaled by a factor of $2$. Thus, the unitary bound $\frac{1}{2}$ of those references translates into $\Delta =1$ here.} $\Delta$ of monopole operators \cite{Gaiotto:2008ak}: a theory is called \emph{good} if all monopole operators have $\Delta >1$; \emph{ugly} if all monopole operators satisfy $\Delta\geq1$, but there exists an operator saturating the unitarity bound. A theory is denoted as \emph{bad} if there exists a monopole operator with $\Delta <1$ violating the unitary bound.

Restricting the focus on \emph{good} 3d $\Ncal=4$ theories has several advantages: (i) No operators decouple during the RG-flow into the IR, and the UV R-symmetry is the same as the IR R-symmetry. (ii) The Coulomb branch is a conical singularity. (iii) Computational tools, like Hilbert series or indices, are convergent.

In addition to the \emph{good, ugly, bad} trichotomy,  \cite{Gaiotto:2008ak} also introduced the notion of \emph{balance}. This condition proved vital for the study of symmetry enhancement due to monopole operators that have the correct R-charges to sit in the multiplet of the symmetry current. In the cases considered --- linear quivers with bifundamental hypermultiplets --- several exact statements had been proven; for example: a linear quiver theory is a \emph{good} if each individual gauge node is \emph{good}. In addition, the \emph{balanced} case coincided with the lower bound (on number of matter fields) for a theory to be still \emph{good}. Thus, it has evolved into a standard argument to compute the \emph{balance} of each node in a quiver theory and infer whether the theory is good or not.

However, as demonstrated in this work, the notion of balance is not readily applicable to Abelian quiver theories with arbitrary charges\footnote{As shown in several examples below, an Abelian quiver that seems to satisfy the \emph{balance} condition may still be not \emph{good} in the strict sense. Hence, one should not equate these two notions.}. Thus, the initial definition \emph{good} in terms of all monopole operators are above the unitarity bound is employed here.

In view of the arguments employed below, another (perhaps indirect) perspective is useful. Recall the basic 3d $\Ncal=4$ mirror duality between a $\urm(1)$ gauge theory with one hypermultiplet of charge one (sometimes called $\mathrm{SQED}_1$) and a free (twisted) hypermultiplet. The $\mathrm{SQED}_1$ is \emph{ugly} as the fundamental monopole operator saturates the unitarity bound and, hence, decouples entirely, which renders the theory free. Below, the mirror theory is utilized and whenever there are free hypermultiplets in the mirror, this serves as a ``smoking gun'' signal for the original theory being \emph{not good}.  

\paragraph{Decay and fission.} The strategy for finding magnetic quivers that produce ICSSs --- denoted as \emph{stable quivers} --- is greatly streamlined by the decay and fission algorithm \cite{Bourget:2023dkj,Bourget:2024mgn}. Using the algorithm, one can construct the putative stratification of the Coulomb branch in question. With the help of additional established tools, one then needs to derive the constraints that reduce this putative stratification to only two symplectic leaves. In Sections~\ref{Sec:Abelian_Chain} and \ref{Sec:Abelian_Cycle} such constraints are derived precisely. 

The algorithm of \cite{Bourget:2023dkj,Bourget:2024mgn} is first applicable to non-simply laced unitary quivers and the stratification is derived by finding all good quiver in the (finite) set of all putative decay and fission products. Here, this algorithm is implicitly extended by refining what constitutes a good Abelian quiver theory with general charges.

\subsection{How to compute the mirror dual?}
\label{Sec:Mirror_Construction}
For the analysis of 3d $\Ncal=4$ Abelian unitary quiver theories with minimal Coulomb branch, a crucial tool is the explicit construction of the mirror dual theory. Specifically, relevant here are the mirrors of quiver theories with just continuous gauge groups composed of $\urm(1)$ factors, but their mirror may or may not have discrete gauge group factors.
The mirror construction is realized by using (the Pontryagin dual of) the cokernel of the charge matrix.
In the remainder of this section, this construction principle is explained and illustrated using three examples.

Consider an arbitrary 3d $\Ncal=4$ Abelian unitary quiver theory, $n_{H}$-many $\Ncal=4$ hypermultiplets and $n_{G}$-many gauge vertices, i.e.\ $\urm(1)$ gauge factors. The charge matrix $\rho$ can be interpreted as a linear map \eqref{eq:charge_matrix_def}.
The cokernel of this map is the finitely generated Abelian group $\Z^{n_{H}}/\mathrm{Im}\left(\rho\right)$. It comes equipped with a surjective morphism $\Z^{n_H} \to \Z^{n_{H}}/\mathrm{Im}\left(\rho\right)$. Applying Pontryagin duality, one finds the dual gauge group $G^\vee = \left(\Z^{n_{H}}/\mathrm{Im}\left(\rho\right)\right)^{\vee}$ and a representation on the matter multiplets $G^\vee \to \urm(1)^{n_H}$.  Concretely, denote by $\left\{\mathcal{X}_{i}\right\}_{i=1}^{n_{H}}$ and $\left\{\mathcal{Y}_{j}\right\}_{j=1}^{n_{G}}$ the generating elements of the co-domain and domain of the charge matrix $\rho$, respectively. The Abelian quotient group $\Z^{n_{H}}/\mathrm{Im}\left(\rho\right)$ is generated by the $\left\{\mathcal{X}_{i}\right\}_{i=1}^{n_{H}}$, with relations between them following from the action of $\rho$ on the set $\left\{\mathcal{Y}_{j}\right\}_{j=1}^{n_{G}}$. To each generating element $\mathcal{X}_{i}$ of the domain one can associate a 3d $\Ncal=4$ hypermultiplet. Solving the relations that define the quotient and taking the Pontryagin dual then yields the gauge group of the mirror dual, as well as the charges of the hypermultiplets and their assignment in flavor groups.

\begin{figure}[t!]
\centering
\begin{subfigure}[t]{0.495\textwidth}
\centering
\includegraphics[page=1]{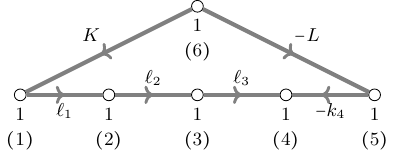}
\caption{}
\label{Fig:Mirror_Example1_6Nodes}
\end{subfigure}
\begin{subfigure}[t]{0.495\textwidth}
\centering
\includegraphics[page=2]{Abelian_Decay_Fission_Fig18.pdf}
\caption{}
\label{Fig:Mirror_Example1_Dual_6Nodes}
\end{subfigure}
\caption{A 3d $\Ncal=4$ cycle quiver theory and its mirror dual theory. \subref{Fig:Mirror_Example1_6Nodes}: In the cycle quiver with $6$ vertices, each gauge vertex carries two descriptors; its rank $1$ and its label in brackets. Each non-simply laced edge denotes the higher gauge charge with respect to the $\urm(1)$ gauge factor positioned at the gauge vertex to which the higher charge is orientated away from. The two charges $K$ and $L$ are defined to be: $K=k_{4}$ and $L=\ell_{1}\ell_{2}\ell_{3}$.  \subref{Fig:Mirror_Example1_Dual_6Nodes}: its mirror dual theory: A SQED-like theory with $6$  pairs of $\Ncal=2$ chiral multiplets with different charges.}
\label{Fig:Mirror_Example_Full}
\end{figure}

\paragraph{Example 1.}
For the first example, consider the 3d $\Ncal=4$ unitary Abelian cycle theory in Figure~\ref{Fig:Mirror_Example1_6Nodes}. 
The charge matrix $\rho:\Z^{6}\to\Z^{6}$ reads
\begin{equation}
\rho =
\begin{pmatrix}
\ell_{1} & -1 & 0 & 0 & 0 & 0 \\
0 & \ell_{2} & -1 & 0 & 0 & 0 \\
0 & 0 & \ell_{3} & -1 & 0 & 0 \\
0 & 0 & 0 & 1 & -k_{4} & 0 \\
0 & 0 & 0 & 0 & 1 & -L \\
-1 & 0 & 0 & 0 & 0 & K 
\end{pmatrix}
\; . \label{Eq:NChain_Charge_6Nodes}
\end{equation}
The cokernel $\Z^{6}/\mathrm{Im}\left(\rho\right)$ for this charge matrix is generated by elements $\left\{\mathcal{X}_{i}\right\}_{i=1}^{6}$ subject to the following relations: 
\begin{align}
    \ell_{1}\mathcal{X}_{1}=\mathcal{X}_{6}
    \,, \quad \mathcal{X}_{1}=\ell_{2}\mathcal{X}_{2}
    \,, \quad \mathcal{X}_{2}=\ell_{3}\mathcal{X}_{3}
    \,, \quad 
    \mathcal{X}_{3}=\mathcal{X}_{4}
    \,, \quad k_{4}\mathcal{X}_{4}=\mathcal{X}_{5}
    \,, \quad 
    L\mathcal{X}_{5}=K\mathcal{X}_{6}
    \,.
\end{align}
Using five of these relations, all the generators can be expressed in terms of $\mathcal X_3$. The remaining relation is trivially satisfied. The cokernel is therefore isomorphic to $\mathbb Z$, generated by $\mathcal X_3$. The (Pontryagin) dual gauge group is $\urm(1)$ and the relations $\mathcal X_1 = \ell_2\ell_3 \mathcal X_3$, $\mathcal X_2 = \ell_3 \mathcal X_3$, etc. give us the charge of the corresponding hypermultiplets:
\begin{align}
\left[\mathcal{X}_{1}\right]=\ell_{2}\ell_{3}
\,, \quad 
\left[\mathcal{X}_{2}\right]=\ell_{3}
\,, \quad 
\left[\mathcal{X}_{3}\right]=\left[\mathcal{X}_{4}\right]=1
\,, \quad 
\left[\mathcal{X}_{5}\right]=k_{4}
\,, \quad 
\left[\mathcal{X}_{6}\right]=\ell_{1}\ell_{2}\ell_{3}
    \,.
\end{align}
Thus, this construction gives the mirror dual shown in Figure~\ref{Fig:Mirror_Example1_Dual_6Nodes}. 
As a remark, this mirror agrees with proposed mirror\footnote{Therein, the kernel of the transposed charge matrix $\mathrm{Ker}\left(\rho^{\mathrm{T}}\right)=\left\{\lambda \,
    \begin{pmatrix}
\ell_{2}\ell_{3}&\ell_{3}&1&1&k_{4}&\ell_{1}\ell_{2}\ell_{3}
    \end{pmatrix}^{\mathrm{T}} \, | \, \lambda\in\Z\right\}$ is utilized to derive the charge matrix of the mirror theory.} of \cite{deBoer:1996ck,Kapustin:1999ha} and reviewed in \cite{Tong:2000ky}, as well as subsequent works.

\begin{figure}[t!]
\centering
\begin{subfigure}[t]{0.495\textwidth}
\centering
\includegraphics[page=1]{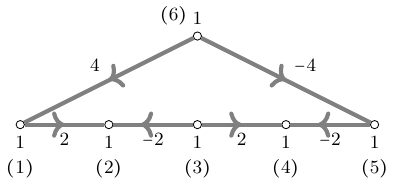}
\caption{}
\label{Fig:Mirror_Example2}
\end{subfigure}
\begin{subfigure}[t]{0.495\textwidth}
\centering
\includegraphics[page=2]{Abelian_Decay_Fission_Fig19.pdf}
\caption{}
\label{Fig:Mirror_Example2_Dual}
\end{subfigure}
\caption{A 3d $\Ncal=4$ Abelian unitary quiver with six vertices (Figure \subref{Fig:Mirror_Example2}) and its mirror dual theory. The mirror dual theory in Figure \subref{Fig:Mirror_Example2_Dual} is equipped with the gauge group $\urm(1)\times\Z_{2}$ and six pairs of $\Ncal=2$ chiral multiplets. Two of them carry charge with respect to both gauge factors in the product group}
\label{Fig:Mirror_Example2_Full}
\end{figure}
\paragraph{Example 2.}
Consider next the quiver theory in Figure~\ref{Fig:Mirror_Example2}. As for the previous example, start with the charge matrix $\rho:\Z^{6}\to\Z^{6}$:
\begin{align}
    \rho=    
    \begin{pmatrix}
        2 & -1 & 0 & 0 & 0 & 0\\
        0 & 1 & -2 & 0 & 0 & 0\\
        0 & 0 & 2 & -1 & 0 & 0\\
        0 & 0 & 0 & 1 & -2 & 0\\
        0 & 0 & 0 & 0 & 1 & -4\\
        -1 & 0 & 0 & 0 & 0 & 4\\
    \end{pmatrix} \; . \label{Eq:Charge_Matrix_Example2_Mirror}
\end{align}
Consider the theory defined by the cokernel $\Z^{6}/\mathrm{Im}\left(\rho\right)$. From the charge matrix \eqref{Eq:Charge_Matrix_Example2_Mirror} follow the relations for the generating elements $\left\{\mathcal{X}_{i}\right\}_{i=1}^{6}$:
\begin{align}
    2\mathcal{X}_{1}=\mathcal{X}_{6}
    \,, \quad 
    \mathcal{X}_{1}=\mathcal{X}_{2}
    \,, \quad 
    2\mathcal{X}_{2}=2\mathcal{X}_{3}
    \,, \quad 
    \mathcal{X}_{3}=\mathcal{X}_{4}
    \,, \quad 
    2\mathcal{X}_{4}=\mathcal{X}_{5}
    \, \quad 
    4\mathcal{X}_{5}=4\mathcal{X}_{6} \; . \label{Eq:Cokernel_Relation_Example2_Mirror}
\end{align}
As above, we can simplify these relations and find an explicit form for the mirror by choosing the right generators. Here, we can choose $   \mathcal V \equiv \mathcal X_1$ and $\mathcal W\equiv \mathcal X_3 -\mathcal X_1 $.
The relations \eqref{Eq:Cokernel_Relation_Example2_Mirror} allow us to express $\mathcal X_1,\ldots \mathcal, X_6$ in terms of $\mathcal V$ and $\mathcal W$, and find the relations that these new generators satisfy:
\begin{subequations}
\begin{gather}
    \mathcal X_1= \mathcal X_2 = \mathcal V, \quad \mathcal X_3 = \mathcal V + \mathcal W, \quad \mathcal X_4 = \mathcal V - \mathcal W  ,\quad\mathcal X_5 = \mathcal X_6 = 2\mathcal V \,,
    \label{Eq:Charge_assignment_Example2_Mirror} \\
    2 \mathcal W=  0 \,.
\end{gather}
\end{subequations}
We find that the cokernel of $\rho$ is isomorphic to $\mathbb Z \times \mathbb Z_2$, and therefore the gauge group of the mirror is $\urm(1) \times \mathbb Z_2$. We can read how charges are assigned to the six hypermultiplets from \eqref{Eq:Charge_assignment_Example2_Mirror}: 
\begin{subequations}
\begin{align}
    [\mathcal X_i]_{\urm(1)} &= \left\{ \begin{array}{cl} 1 & \text{if } i =1,2,3,4 \\
    2 & \text{if } i=5,6\end{array}\right. \\
    [\mathcal X_i]_{\mathbb Z_2} &=\left\{ \begin{array}{cl} 0 & \text{if } i =1,2,5,6 \\
    1 & \text{if } i=3,4\end{array}\right.
\end{align}
\end{subequations}
This yields the theory as in Figure~\ref{Fig:Mirror_Example2_Dual}. In contrast to the first example, the discrete gauge factor is not captured just computing the kernel\footnote{One finds $\mathrm{Ker}\left(\rho^{\mathrm{T}}\right)=\left\{\lambda \,
    \begin{pmatrix}
        1&1&1&1&2&2
    \end{pmatrix}^{\mathrm{T}} \, | \, \lambda\in\Z\right\}$.} of the transpose of the charge matrix. Instead, the two hypermultiplets that carry non-trivial charge with respect to the $\Z_{2}$ factor would seem to combine with the two hypermultiplets that only carry charge $1$ with respect to the $\urm(1)$ gauge factor to a rank four flavor group with gauge group $\urm(1)$. The correct mirror theory is the one with the discrete gauge factor, constructed using the cokernel.

In fact, the appearance of the extra discrete $\Z_2$ gauge factor in the SQED-like mirror can be understood from the perspective of gauging discrete subgroups and arising 1-form symmetries, along the lines of \cite{Nawata:2023rdx}. Concretely, the charge $2$ of the two hypermultiplets connecting to the 3rd node of the quiver in Figure \ref{Fig:Mirror_Example2} can be understood as result of gauging a $\Z_2$ subgroup of the topological symmetry associated to that node. As discussed, for example, in \cite{Nawata:2023rdx,Bhardwaj:2023zix} this then translates to gauging a discrete $\Z_2$ subgroup of the SQED flavor symmetry (see also \cite{Gaiotto:2014kfa,Hsin:2018vcg,Beratto:2021xmn,Mekareeya:2022spm} for related works).

\paragraph{Example 3.}
\begin{figure}[t!]
\centering
\begin{subfigure}[t]{1\textwidth}
\centering
\includegraphics[page=1]{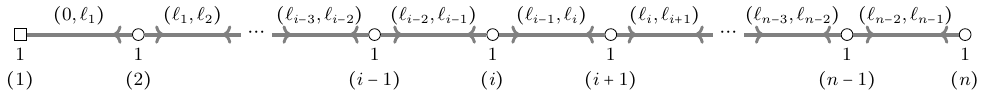}
\caption{}
\label{Fig:Mirror_Construction_Example3}
\end{subfigure}
\begin{subfigure}[t]{1\textwidth}
\centering
\includegraphics[page=2]{Abelian_Decay_Fission_Fig21.pdf}
\caption{}
\label{Fig:Mirror_Construction_Example3_Mirror}
\end{subfigure}
\caption{\subref{Fig:Mirror_Construction_Example3}: Example of a 3d $\Ncal=4$ quiver whose mirror theory is, in general, not a quiver and only contains discrete gauge groups. \subref{Fig:Mirror_Construction_Example3_Mirror}: In the special case of $\ell_i=1$ for all $i \neq k,n-1$, the mirror admits a quiver description.}
\label{Fig:Mirror_Example3_Full}
\end{figure}
Consider the quiver in Figure \ref{Fig:Mirror_Construction_Example3}. Using the cokernel method, one computes that the mirror theory can be described as $n-1$ hypermultiplets that are charged under $n-1$ many discrete cyclic groups as follows: 
\begin{center}
    \begin{tabular}{c|ccccc}
    Group&$\mathcal{X}_{1}$&$\mathcal{X}_{2}$&$\cdots$&$\mathcal{X}_{n-2}$&$\mathcal{X}_{n-1}$\\
    \hline
    $\Z_{\ell_{1}}$&1&0&$\cdots$&0&0 \\
    $\Z_{\ell_{2}}$&1&1&$\cdots$&0&0 \\
    $\vdots$&&&&& \\
    $\Z_{\ell_{n-2}}$&1&1&$\cdots$&1&0 \\
    $\Z_{\ell_{n-1}}$&1&1&$\cdots$&1&1 \\
    \end{tabular}
\end{center}
In general, this is not a ``quiver'' theory. If $\ell_i = 1$ for all $i$ except for two indices, say $i=k$ and the last one $i=n$, then the mirror is the quiver shown in Figure \ref{Fig:Mirror_Construction_Example3_Mirror}. 

\paragraph{Remarks.}
It is worth noting that the choice of the generators and the calculation of the mirror gauge group can be automated, using the Smith normal form of the charge matrix. Starting with the generators $\{\mathcal X_i\}_{i=1}^{n_H}$, one for each hypermultiplet, and the relations $\sum_{i=1}^{n_H} \rho_{i \alpha}\mathcal X_i = 0$, it computes a minimal set of linearly independent relations, and a change of basis on the generators which diagonalizes these relations. Concretely, we get a set of generators $\{\mathcal V_j\}$, such that the relations reduce to $d_j \mathcal V_j = 0$. The cokernel is then the product of a copy of $\mathbb Z$ for each $\mathcal V_j$ with $d_j=0$ and a cyclic group of order $d_j$ for each $\mathcal V_j$ with $d_j>1$. The former yield $\urm(1)$ factors in the mirror gauge group, while the latter give discrete factors. By inverting the change of basis and expressing the original generators $\mathcal X_i$ in terms of the $\mathcal V_j$, we find the charges of the hypermultiplets in the mirror.

It can be shown that the connected part of the gauge group constructed using the cokernel of the charge matrix is the same as the gauge group constructed from the kernel of its transpose. Furthermore, the charges of the hypermultiplets in the mirror theory agree in both constructions. The only difference is that the gauge group obtained using the cokernel can contain a discrete factor, which can be identified with the discrete part of the kernel of the representation morphism $\urm(1)^{n_G}\to \urm(1)^{n_H}$. This matches the results of \cite{deBoer:1996ck} where the kernel was used to construct duals of unitary Abelian theories. Computing explicitly the Coulomb and Higgs branch metrics in the original and mirror theories, a mismatch was found which corresponds exactly to the additional discrete factor provided by the cokernel.

\subsection{Which quivers define the same symplectic singularity?}
\label{Sec:Quiver_reduced_form}
The 3d $\Ncal=4$ Coulomb branch, as considered via various approaches \cite{Cremonesi:2013lqa,Bullimore:2015lsa,Nakajima:2015txa,Braverman:2016wma}, and the mirror dual theory, given by the cokernel construction described above\footnote{Strictly speaking, the gauge group and the representation on hypermultiplets in the dual theory is unchanged.}, only depend on the image lattice of the charge matrix. Therefore, two quivers whose charge matrices $\rho_{1}$ and $\rho_{2}$ are related by $\rho_{1} = \rho_{2}\, U$ with $U \in\operatorname{GL}(n,\Z)$ have the same mirror dual and the same Coulomb branch\footnote{In general, if $\rho$ is the charge matrix of a quiver and $U$ a unimodular matrix, $\rho \,U$ does not correspond to a quiver theory, as some hypermultiplets are charged under more than two gauge group factors. Moreover, this transformation typically changes the kinetic term for the gauge fields. This, however, does not affect the Coulomb branch geometry.}. On top of this, when ungauging the connected kernel of the gauge group, one can add to or remove from the charge matrix a column which is an integer linear combination of the others. 

This freedom of reparametrizing the theory can be used to put a quiver theory into a simpler form. In the following, two examples are considered.

\paragraph{Example 1.}
\begin{figure}[t!]
\begin{subfigure}[t]{0.5\textwidth}
    \centering
    \includegraphics[page=2]{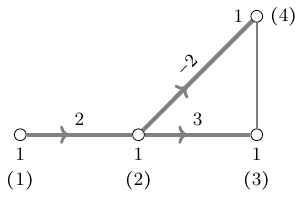}
\caption{}
\label{Fig:Extended_Cycle_Example_1_ReducedExample1}
\end{subfigure}
\begin{subfigure}[t]{0.5\textwidth}
    \centering
    \includegraphics[page=4]{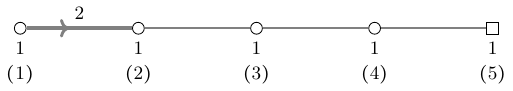}
\caption{}
\label{Fig:Extended_Cycle_Example_2_ReducedExample1}
\end{subfigure}
\caption{An Abelian extended cyclic quiver (Figure \subref{Fig:Extended_Cycle_Example_1_ReducedExample1}) and an Abelian chain quiver (Figure \subref{Fig:Extended_Cycle_Example_2_ReducedExample1}) that are related to each other by reparametrization.}
\label{Fig:Extended_Cycle_Example_ReducedExample1}
\end{figure}
For the first example, consider the extended cyclic quiver in Figure \ref{Fig:Extended_Cycle_Example_1_ReducedExample1}, defined by the charge matrix\footnote{Note that this quiver does not have a well-defined length in the sense of Section \ref{Sec:Abelian_Cycle_Extension}.}
\begin{align}
    \rho=
    \begin{pmatrix}
        2 & -1 & 0 & 0\\
        0 & 3 & -1 & 0\\
        0 & 0 & 1 & -1\\
        0 & -2 & 0 & 1\\
    \end{pmatrix} \; .
\end{align}
Applying $U\in\mathrm{GL}\left(4,\Z\right)$ yields
\begin{align}
    U\equiv
    \begin{pmatrix}
        1 & 0 & 0 & 0\\
        0 & 1 & 0 & 0\\
        0 & 2 & 1 & 0\\
        0 & 2 & 0 & 1\\
    \end{pmatrix}\, , \quad
    \rho\, U =
    \begin{pmatrix}
        2 & -1 & 0 & 0\\
        0 & 1 & -1 & 0\\
        0 & 0 & 1 & -1\\
        0 & 0 & 0 & 1\\
    \end{pmatrix} \; ,
\end{align}
with $\rho\, U$ now defining the charge matrix of the chain quiver in Figure \ref{Fig:Extended_Cycle_Example_2_ReducedExample1}. Note that the reparametrization resulted in a framed quiver, which is expressed in the charge matrix via the fourth row encoding a hypermultiplet that carries gauge charge with respect to only one $\urm(1)$ gauge factor.

\paragraph{Example 2.} Next, consider the chain quiver in Figure \ref{Fig:Complete_Class_AbelianChain_Minimal} of Section \ref{Sec:Abelian_Chain_Equivalence}. The charge matrix reads
\begin{equation}
\rho=
\begin{tikzpicture}[baseline]
\matrix (m) [matrix of math nodes,nodes in empty cells,right delimiter={)},left delimiter={(} ]{
\ell_1 & -k_1 & 0 & 0 & & & & & 0  \\
0 & \ell_2 & -k_2 & 0 & & & & & 0  \\
0 & 0 & \ell_3 & -k_3 & & & & & 0  \\
& & & & & & & &  \\
& & & & & & & & \\
& & & & & & & & \\
0 & & & & & & \ell_{n-2} & -k_{n-2} & 0 \\
0 & & & & & & 0 & \ell_{n-1} & -k_{n-1} \\
} ;
\draw[loosely dotted, thick] (m-1-4)-- (m-1-9);
\draw[loosely dotted, thick] (m-3-9)-- (m-7-9);
\draw[loosely dotted, thick] (m-3-1)-- (m-7-1);
\draw[loosely dotted, thick] (m-8-1)-- (m-8-7);
\draw[loosely dotted, thick] (m-4-4)-- (m-7-7);
\end{tikzpicture} \; . \label{Eq:NChain_Charge_Example}
\end{equation} 
For $1\leq i\leq n-1$, let $\delta_i \equiv\gcd\left(k_i,\prod_{j=1}^i \ell_j\right)$. This $\gcd$-expression can be decomposed via Bézout coefficients\footnote{Note that this decomposition is not unique.}, i.e.\ integer-valued coefficients $u_i,v_i$, such that $\delta_i = u_i \left(\prod_{j=1}^i \ell_j\right) - v_i k_i$. Using this, one can define for all $1\leq i\leq n-1$ the following set of $n\times n$ matrices:
\begin{equation}
U_i \equiv \left(\begin{array}{c|cc|c}
I_{i-1} &&\\ \hline
& u_{i}\prod_{j=1}^{i-1}\delta_{j} & \frac{k_i}{\delta_i} \\
& v_i & \prod_{j=1}^{i}\left(\frac{\ell_{j}}{\delta_{j}}\right) \\ \hline 
& & & I_{n-i-1}
\end{array}\right)\in\operatorname{GL}(n,\mathbb Z) \;.
\end{equation}
With this set of matrices, one can now define the \emph{reduced} quiver representation for the chain quiver in Figure \ref{Fig:Complete_Class_AbelianChain_Minimal}:
\begin{equation}
\rho\, U_{1}\,U_{2}\cdots U_{n-1}=
\begin{tikzpicture}[baseline]
\matrix (m) [matrix of math nodes,nodes in empty cells,right delimiter={)},left delimiter={(} ]{
\delta_1 & 0 & 0 & & & & & & 0\\
q_2 & \delta_2 & 0 & & & & & & 0\\
0 & q_3 & \delta_3 & & & & & & 0\\
& & & & & & & & \\
& & & & & & & & \\
& & & & & & & & \\
0 & & & & & q_{n-2} & \delta_{n-2} & 0 & 0 \\
0 & & & & & 0 & q_{n-1} & \delta_{n-1} & 0\\
} ;
\draw[loosely dotted, thick] (m-1-3)-- (m-1-9);
\draw[loosely dotted, thick] (m-3-9)-- (m-7-9);
\draw[loosely dotted, thick] (m-3-1)-- (m-7-1);
\draw[loosely dotted, thick] (m-8-1)-- (m-8-6);
\draw[loosely dotted, thick] (m-3-3)-- (m-7-7);
\node at (0,-3) {with \quad $q_{i}\equiv \ell_{i}\upsilon_{i-1} \quad \forall\, i:\, 2\leq i\leq n-1$ \; .};
\end{tikzpicture} \label{Eq:NChain_Charge_Reduced_Example}
\end{equation}
Notice that the last column is zero; the corresponding $\urm(1)$ factor of the gauge group acts trivially and is removed from the gauge action. That is to say, this reparametrization results in a framed quiver for which the first row in \eqref{Eq:NChain_Charge_Reduced_Example} defines the only hypermultiplet that is charged with respect to the associated flavor group of rank $1$, analogue to the first example.

There are some technical subtleties that need to be remarked on: For one, note that the magnetic fluxes for the framed reduced chain quiver defined by the charge matrix \eqref{Eq:NChain_Charge_Reduced_Example} are integer-valued, i.e.\ the co-character lattice for the overall gauge group is not modified\footnote{When ungauging (the connected part of) the kernel of the gauge action, one has a choice of representatives of the resulting equivalence classes. The usual convention (when possible) is a choice such that one of the $\urm(1)$ gauge factors decouples completely, similar as for the reduced quiver form defined by the charge matrix \eqref{Eq:NChain_Charge_Reduced_Example}. Generically, for the choice made, it might be necessary to modify the co-character lattice.}. That is no trivial statement, since framing the chain quiver defined by the charge matrix \eqref{Eq:NChain_Charge_Example} for generic values of the charges $\ell_{i}$ and $k_{i}$ might require a modification of its co-character lattice. Furthermore, when unframing the reduced chain quiver defined by \eqref{Eq:NChain_Charge_Reduced_Example}, one cannot equip the hypermultiplet connected to the associated $\urm(1)$ gauge factor with an arbitrary charge. Rather, one has to take into account the $\gcd$-relations encoded in the parametrization of the reduced form (see for example Section \ref{Sec:AbelianChain_Remarks_Redundancy}), as well as the integer-valued co-character lattice\footnote{Generally, when unframing a quiver theory, the data regarding a modification of the co-character lattice is required. Otherwise, one needs to make a choice when introducing charges for the hypermultiplets.}.

\section{Tree-like quivers with minimal Coulomb branch}
\label{Sec:Abelian_Chain}
In this section the complete class of 3d $\Ncal=4$ Abelian unitary $n$-vertices tree-like quiver theories with $\left(\ell,-k\right)$-edges whose Coulomb branch is an isolated conical symplectic singularity (ICSS) is defined and their Coulomb branch geometries are identified.
The complete class of Abelian unitary $n$-vertex chain quiver theories with $\left(\ell,-k\right)$-edges --- also referred to as \emph{chain quivers} --- with minimal Coulomb branch is derived using two complementary approaches in Section~\ref{Sec:MinimalAbelianChain}. Following that, in Section~\ref{Sec:AbelianTreeExtension} it is shown that this class of chain quivers is the only class of Abelian unitary tree-like quiver theories whose Coulomb branch is an ICSS.

\subsection{Chain quivers}
\label{Sec:MinimalAbelianChain}

\begin{figure}[t!]
\centering
\includegraphics[page=1]{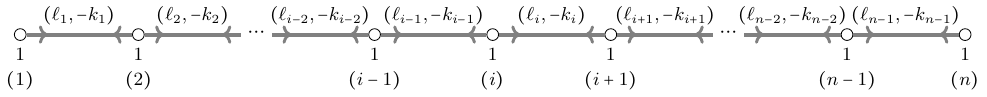}
\caption{Generic 3d $\Ncal=4$ Abelian unitary $n$-vertices chain quiver theory with $\left(\ell,-k\right)$-edges. Each vertex carries two descriptors: $1$ denoting the rank of the associated $\urm(1)$ gauge group and in brackets the vertex label. The $\left(n-1\right)$-many edges are equipped with the integers $\ell_{i}$, orientated away from vertex $\left(i\right)$, and $-k_{i}$, orientated towards vertex $\left(i\right)$. The factor $\ell_{i}$ is the (higher) $\urm(1)_{(i)}$ gauge charge of the hypermultiplet connecting vertices $\left(i\right)$ and $\left(i+1\right)$ and, analogously, the factor $-k_{i}$ denotes the $\urm(1)_{\left(i+1\right)}$ gauge charge. This chain quiver exhibits a minimal Coulomb branch if and only if the set of constraints \eqref{Eq:AbelianChain_GCD_Condition} for the gauge charges are fulfilled.}
\label{Fig:Complete_Class_AbelianChain_Minimal}
\end{figure}
The first class of 3d $\Ncal=4$ Abelian unitary tree-like quivers that is considered are chain quivers as shown in Figure \ref{Fig:Complete_Class_AbelianChain_Minimal}. As is shown below, a quiver theory of this form exhibits a minimal Coulomb branch if and only if the following set of constraints is fulfilled:
\begin{subequations}
\label{Eq:AbelianChain_GCD_Condition}
\begin{alignat}{2}
    & \gcd\left(\ell_{1},k_{j}\right)=1 & \qquad &\forall \, j: \, 1\leq j\leq n-2 \; , \label{Eq:AbelianChain_GCD_Condition1}
    \\
    & \gcd\left(\ell_{i},k_{n-1}\right)=1 & \qquad &\forall \, i: \, 2\leq i\leq n-1 \; , \label{Eq:AbelianChain_GCD_Condition2}
    \\
    & \gcd\left(\ell_{i},k_{j}\right)=1 & \qquad &\forall \, i,j: \, 2\leq i\leq j\leq n-2 \; , \label{Eq:AbelianChain_GCD_Condition3}
    \\
    & \gcd\left(\ell_{1},k_{n-1}\right)>1 \;.  & &  \label{Eq:AbelianChain_GCD_Condition4}
\end{alignat}
\end{subequations}
Note that due to the constraint \eqref{Eq:AbelianChain_GCD_Condition4} all the chain quivers with minimal Coulomb branch are classified as good theories according to the usual balance-based definition \cite{Gaiotto:2008ak}. However, as anticipated in Section~\ref{sec:good-balance}, this definition is not suitable for Abelian unitary theories with arbitrary charges. The conformal dimension of monopole operators yields an unambiguous definition.

For a more intuitive interpretation of the set of constraints \eqref{Eq:AbelianChain_GCD_Condition}, from the point of view of the stratification of the Coulomb branch, the conditions \eqref{Eq:AbelianChain_GCD_Condition1}, \eqref{Eq:AbelianChain_GCD_Condition2} and \eqref{Eq:AbelianChain_GCD_Condition3} effectively exclude all possible decay and fission-channels of the initial chain quiver with $n$-many vertices to any embedded chain quiver with $m$-many vertices $\left(2\leq m\leq n-1\right)$ or products of such chain quivers\footnote{Note the subtlety: The decay and fission algorithm relies on good sub-quivers. Enforcing  \eqref{Eq:AbelianChain_GCD_Condition1} -- \eqref{Eq:AbelianChain_GCD_Condition3} only allows for sub-quivers that are \emph{not good} in both the usual balance condition as well as in the physical good condition.}.

The geometry of the minimal Coulomb branch of a chain quiver in Figure~\ref{Fig:Complete_Class_AbelianChain_Minimal}, defined by the constraints \eqref{Eq:AbelianChain_GCD_Condition}, is identified as an orbifold
\begin{subequations}
\begin{align}
    \mathcal{M}_{\Coulomb} \cong \left(\mathbb{H}^{n-1}/\Z_{\gcd\left(\ell_{1},k_{n-1}\right)}   \right)^{[\sigma]}
    \stackrel{\eqref{eq:h-def}}{\equiv} 
    h_{n-1,\gcd\left(\ell_{1},k_{n-1}\right),\sigma}
    \; , 
    \label{eq:minimalChain_CB_geometry}
\end{align}
with $\Z_{\gcd\left(\ell_{1},k_{n-1}\right)}$ the orbifold group and $\sigma$ denoting the charge vector of the $\left(n-1\right)$-many $\mathbb{H}$ factors.

Utilizing 3d $\Ncal=4$ mirror symmetry, this Coulomb branch has a Higgs branch realization by interpreting the orbifold group as a discrete gauge group and the $\mathbb{H}$ factors as hypermultiplets. Then expressing the 
hypermultiplets as $\left(n-1\right)$-many pairs of 3d $\Ncal=2$ chiral multiplets $(\mathcal{A}_{i},\widetilde{\mathcal{A}_{i}})$, the gauge charges are defined as follows:
\begin{alignat}{2}
   \left[\mathcal{A}_{1}\right]&=\prod_{m=2}^{n-1}L_{m} \, , \quad \left[\mathcal{A}_{n-1}\right]=\prod_{n=1}^{n-2}K_{n} \,
   \\
   \left[\mathcal{A}_{i}\right]&=\prod_{m=i+1}^{n-1}L_{m}\prod_{n=1}^{i-1}K_{n} & \qquad &\forall \, i: \, 2\leq i\leq n-2 \; ,
    \\
  \text{with} \qquad   K_{i}&\equiv \frac{k_{1}}{g_{i}} & \qquad &\forall \, i: \, 1\leq i \leq n-2 \; ,
    \\
     L_{j}&\equiv \frac{\ell_{j}}{g_{j-1}} &\qquad &\forall \, j: \, 2\leq j \leq n-1 \; ,
    \\
    g_{p}&\equiv \gcd\left(\ell_{p+1},k_{p}\right) &\qquad &\forall \, p: \, 1\leq p \leq n-1 \; .
\end{alignat}
\end{subequations}
Below, this result is derived in two different ways. First, the complete class of chain quivers with minimal Coulomb branch is explicitly constructed via an induction in Section~\ref{Sec:Abelian_Chain_Induction}. Afterwards, in Section~\ref{Sec:Abelian_Chain_Equivalence}, the equivalence between the statements of having a chain quiver with minimal Coulomb branch and having a chain quiver that fulfills the constraints \eqref{Eq:AbelianChain_GCD_Condition} is proven directly. The reason for presenting both derivations is as follows: The inductive approach offers insight into how the constraints \eqref{Eq:AbelianChain_GCD_Condition} arise by applying the decay and fission algorithm. The latter derivation, on the other hand, offers less insight into the underlying physical process, but does not rely on the decay and fission algorithm. Therefore, it can be viewed as a consistency check of the algorithm. Furthermore, in contrast to the approach by induction, this derivation is done in the reduced chain quiver representation introduced in Section~\ref{Sec:Quiver_reduced_form}, showcasing how the constraints \eqref{Eq:AbelianChain_GCD_Condition} manifest in both distinct representations.

\subsubsection{Constructing stable Abelian chain quivers via induction}
\label{Sec:Abelian_Chain_Induction}
For the following induction it is necessary to (schematically) construct the 3d $\Ncal=4$ mirror dual theory for a generic chain quiver with $n$-many vertices as in Figure~\ref{Fig:Complete_Class_AbelianChain_Minimal} by applying the prescription discussed in the Section~\ref{Sec:Mirror_Construction}. The charge matrix $\rho: \Z^{n} \rightarrow \Z^{n-1}$ in this case reads 
\begin{equation}
\rho=
\begin{tikzpicture}[baseline]
\matrix (m) [matrix of math nodes,nodes in empty cells,right delimiter={)},left delimiter={(} ]{
\ell_1 & -k_1 & 0 & 0 & & & & & 0  \\
0 & \ell_2 & -k_2 & 0 & & & & & 0  \\
0 & 0 & \ell_3 & -k_3 & & & & & 0  \\
& & & & & & & &  \\
& & & & & & & & \\
& & & & & & & & \\
0 & & & & & & \ell_{n-2} & -k_{n-2} & 0 \\
0 & & & & & & 0 & \ell_{n-1} & -k_{n-1} \\
} ;
\draw[loosely dotted, thick] (m-1-4)-- (m-1-9);
\draw[loosely dotted, thick] (m-3-9)-- (m-7-9);
\draw[loosely dotted, thick] (m-3-1)-- (m-7-1);
\draw[loosely dotted, thick] (m-8-1)-- (m-8-7);
\draw[loosely dotted, thick] (m-4-4)-- (m-7-7);
\end{tikzpicture} \; . \label{Eq:NChain_Charge}
\end{equation} 
The data of the gauge group, flavor group and charges of the hypermultiplets of the mirror theory is encoded in the Pontryagin dual of the cokernel $\left(\Z^{n-1}/\mathrm{Im}\left(\rho\right)\right)^{\vee}$ of $\rho$.
Denote by $\left\{\mathcal{X}_{i}\right\}_{i=1}^{n-1}$ the generating elements\footnote{With each generating element $\mathcal{X}_{i}$ one associates a $\Ncal=2$ chiral pair $(\mathcal{A}_{i},\widetilde{\mathcal{A}_{i}})$.} of the Abelian group $\Z^{n-1}/\mathrm{Im}\left(\rho\right)$ which are subject to relations 
\begin{subequations}
\label{Eq:General_AbelianChain_Relations_Mirror}
\begin{align}
     \ell_{1}\mathcal{X}_{1}&=0 \; ,
    \\
     k_{n-1}\mathcal{X}_{n-1}&=0 \; ,
    \\
     k_{i}\mathcal{X}_{i}&=\ell_{i+1}\mathcal{X}_{i+1} \qquad \forall \, i: \, 1\leq i\leq n-2 \; , \label{Eq:General_AbelianChain_Relations_Charge}
\end{align}
\end{subequations}
which follow directly from the action of $\rho$ on the generating elements $\{\mathcal{Y}_{j}\}_{j=1}^{n}$ of the domain $\Z^{n}$. The relations \eqref{Eq:General_AbelianChain_Relations_Mirror} put together yield
\begin{subequations}
\begin{alignat}{2}
    & \prod_{j=1}^{i}\ell_{j}\mathcal{X}_{i}=0 \quad \land \quad \prod_{j=i}^{n-1}k_{j}\mathcal{X}_{i}=0 & \qquad &\forall \, i: \, 2\leq i \leq n-2 \label{Eq:General_AbelianChain_Relation_Mod0}
    \\
    \Leftrightarrow \quad & \frac{\prod_{j=1}^{i}\ell_{j}}{g_{i,i}}\left(g_{i,i}\mathcal{X}_{i}\right)=0 \quad \land \quad \frac{\prod_{j=1}^{n-1}k_{j}}{g_{i,i}}\left(g_{i,i}\mathcal{X}_{i}\right)=0 &\qquad &\forall \, i: \, 2\leq i \leq n-2 \;, \label{Eq:General_AbelianChain_Relations_Mod1} \\
 &\quad \text{with}  \qquad   g_{p,q}\equiv\gcd\left(\prod_{j=1}^{p}\ell_{j},\prod_{m=q}^{n-1}k_{m}\right) \; . \label{Eq:GeneralAbelianChain_GCD_Function}
\end{alignat}
\end{subequations}
Since the factors $\left(\prod_{j=1}^{i}\ell_{j}\Big/g_{i,i}\right)$ and $\left(\prod_{j=1}^{n-1}k_{j}\Big/g_{i,i}\right)$ in \eqref{Eq:General_AbelianChain_Relations_Mod1} do not share a common divisor by definition of \eqref{Eq:GeneralAbelianChain_GCD_Function}, the order of the element $g_{i,i}\mathcal{X}_{i}$ is one, i.e. 
\begin{align}
    g_{i,i}\mathcal{X}_{i}=0 \; . \label{Eq:General_AbelianChain_Relations_GCD}
\end{align}
Note that the constraint \eqref{Eq:General_AbelianChain_Relations_GCD} defines a necessary requirement for the chain quiver to be good: If for some $1\leq i\leq n-1$ the $\gcd$-expression $g_{i,i}=1$, then the hypermultiplet associated with the generating element $\mathcal{X}_{i}$ becomes free and decouples from the theory. Therefore, in order for the chain quiver to be good, it is necessary\footnote{One can construct an example to showcase that this is a not a sufficient condition.} that $g_{i,i}>1$ for all $i$ with $1\leq i\leq n-1$. Comparing this necessary condition with the usual balance condition for a good quiver theory, one finds that chain quivers that are not good in the balance-based sense are not good in the physical good definition as well, since the necessary condition is violated. In particular, one has $g_{1,1}=1$ or $g_{n-1,n-1}=1$. However, as anticipated in Section~\ref{sec:good-balance}, a chain quiver that is good in the usual balance sense may still have monopole operators saturating the unitarity bound, which renders them not good in the physical sense.

Using the (schematic) mirror theory to a generic chain quiver as in Figure \ref{Fig:Complete_Class_AbelianChain_Minimal} one can proceed by induction: For the base case, the chain quivers with two and three vertices are considered. For the two-vertices chain, the constraints defining a theory with minimal Coulomb branch are derived and it is shown that using these constraints one can further define the stable three-vertices chain quivers. For the induction step, one assumes the constraints extended to a chain quiver with $\left(n-1\right)$-many vertices to hold true and, again, shows that enforcing these assumptions implies a minimal Coulomb branch for the chain quiver with $n$-many vertices. 

\paragraph{Abelian chain quivers with two vertices.} 
\begin{figure}[t]
\centering
\begin{subfigure}[t]{0.4\textwidth}
\centering
\includegraphics[page=1]{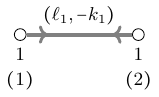}
\caption{}
\label{Fig:2VertexChain_Quiver}
\end{subfigure}
    \hspace{0.3em}
\begin{subfigure}[t]{0.4\textwidth}
\centering
\includegraphics[page=2]{Abelian_Decay_Fission_Fig7.pdf}
\caption{}
\label{Fig:2VertexChain_Mirror}
\end{subfigure}
\caption{Two-vertices chain quiver (\subref{Fig:2VertexChain_Quiver})  and its 3d $\Ncal=4$ mirror dual (\subref{Fig:2VertexChain_Mirror}).  
The mirror dual has a discrete gauge group $\Z_{\gcd\left(\ell_{1},k_{1}\right)}$ and matter content composed of one $\Ncal=2$ chiral-pair $(\mathcal{A},\widetilde{\mathcal{A}})$. 
}
\label{Fig:2VertexChain_QuiverAndMirror}
\end{figure}

Consider the chain quiver with two vertices displayed in Figure~\ref{Fig:2VertexChain_Quiver}. For this quiver theory, the charge matrix reads
\begin{align}
\rho=
    \begin{pmatrix}
        \ell_{1} & -k_{1}
    \end{pmatrix}
\end{align}
and the relations \eqref{Eq:General_AbelianChain_Relations_Charge} and \eqref{Eq:General_AbelianChain_Relations_GCD} defining the cokernel simplify to
\begin{align}
    g_{1,1}\mathcal{X}_{1}=0 \; , \quad g_{1,1}=\gcd\left(\ell_{1},k_{1}\right)
\end{align}
for the single generating element $\mathcal{X}_{1}$. From that, it follows that the orbifold theory in Figure~\ref{Fig:2VertexChain_Mirror} is the 3d $\Ncal=4$ mirror theory of the chain theory.
Note that for $\gcd\left(\ell_{1},k_{1}\right)=1$ the mirror theory, and hence the initial chain quiver, is trivial; the hypermultiplet becomes free and decouples from the theory. If $\gcd\left(\ell_{1},k_{1}\right)>1$, the mirror theory exhibits a minimal Higgs branch since higgsing the (monomial) gauge-invariant operator formed from $(\mathcal{A},\widetilde{\mathcal{A}})$ breaks the discrete gauge group completely. This implies a minimal Coulomb branch for the chain theory as well and, in addition, the geometry is defined by the orbifold
\begin{align}
    \mathcal{M}_{\Coulomb}\cong \left(\mathbb{H}/\Z_{\gcd\left(\ell_{1},k_{1}\right)}  \right)^{[1]}
    \stackrel{\eqref{eq:h-def}}{=}
    h_{1,\gcd\left(\ell_{1},k_{1}\right),1} =
    h_{1,\gcd\left(\ell_{1},k_{1}\right)} = A_{\gcd\left(\ell_{1},k_{1}\right)}
\end{align}
where $[1]$ denotes the charge $1$ with respect to the discrete group.

In conclusion, the following holds for the two-vertices chain theory of Figure \ref{Fig:2VertexChain_Quiver}: 
\begin{align}
      \left( \substack{\text{two-vertices chain}  \\ \text{(Figure~\ref{Fig:2VertexChain_Quiver})}  }\right) =
    \begin{cases}
        \substack{ \text{non-trivial theory} \\ \text{and minimal Coulomb branch $\mathcal{M}_{\Coulomb}$}} & \Leftrightarrow  \gcd\left(\ell_{1},k_{1}\right)>1 \; , \\ 
        \\
        \substack{\text{trivial theory} \\  \text{(and trivial Coulomb branch $\mathcal{M}_{\Coulomb}$)}} & \Leftrightarrow  \gcd\left(\ell_{1},k_{1}\right)=1 \; .
    \end{cases}
\end{align}

\paragraph{Abelian chain quivers with three vertices.}

\begin{figure}[t!]
\centering
\begin{subfigure}[t]{0.3\textwidth}
\centering
\includegraphics[page=1]{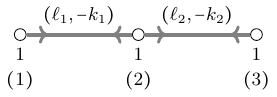}
\caption{}
\label{Fig:3VertexChain_Quiver}
\end{subfigure}
\begin{subfigure}[t]{0.475\textwidth}
\centering
\includegraphics[page=2]{Abelian_Decay_Fission_Fig8.pdf}
\caption{}
\label{Fig:3VertexChain_DecayAndFission}
\end{subfigure}
\begin{subfigure}[t]{0.21\textwidth}
\centering
\includegraphics[page=3]{Abelian_Decay_Fission_Fig8.pdf}
\caption{}
\label{Fig:3VertexChain_Mirror}
\end{subfigure}
\caption{Three-vertices chain quiver (\subref{Fig:3VertexChain_Quiver}) and its generic Coulomb branch Hasse diagram (\subref{Fig:3VertexChain_DecayAndFission}), derived using the decay and fission algorithm. In case the constraints $\gcd\left(\ell_{i},k_{i}\right)=1$ for $i\in\left\{1,2\right\}$ are fulfilled, Figure \subref{Fig:3VertexChain_Mirror} displays the 3d $\Ncal=4$ mirror theory to Figure \subref{Fig:3VertexChain_Quiver}, equipped with the discrete gauge group $\Z_{\gcd\left(\ell_{1},k_{2}\right)}$ and two $\Ncal=2$ chiral-pairs $\{(\mathcal{A}_{i},\widetilde{\mathcal{A}_{i}})\}_{i=1}^{2}$ of gauge charge $L_{2}$ and $K_{1}$ (see \eqref{Eq:AbelianChainExample3Nodes_Charges}), respectively.
}
\label{Fig:3VertexChain_Quiver_DecayAndFission}
\end{figure}

Continue with the chain quiver with three vertices as displayed in Figure \ref{Fig:3VertexChain_Quiver}. Using the decay and fission algorithm, one can deduce the generic Hasse diagram as shown in Figure~\ref{Fig:3VertexChain_DecayAndFission}. In order to minimize this Hasse diagram, one needs to trivialize the two decay-channels. Using the previously gained insight into the chain quiver with two vertices, this means that one needs to enforce the constraints
\begin{align}
    \gcd\left(\ell_{1},k_{1}\right)=1 \quad \land \quad \gcd\left(\ell_{2},k_{2}\right)=1 \; , \label{Eq:3VertexChain_MinimalCondition1}
\end{align}
because the two potential decay products are then trivial. In particular, they are not good and, therefore, not allowed as decay products in the decay and fission algorithm.

The charge matrix $\rho$ of the chain quiver in Figure \ref{Fig:3VertexChain_Quiver} reads
\begin{align}
\rho=
    \begin{pmatrix}
    \ell_1 & -k_1 & 0 \\
    0 & \ell_2 & -k_2
    \end{pmatrix} \label{Eq:3NodeChain_Charge_Matrix}
\end{align}
and the relations \eqref{Eq:General_AbelianChain_Relations_Charge} and \eqref{Eq:General_AbelianChain_Relations_GCD} reduce to
\begin{align}
    g_{1,1}\mathcal{X}_{1}=0 \quad \land \quad g_{2,2}\mathcal{X}_{2}=0 \quad \land \quad k_{1}\mathcal{X}_{1}=\ell_{2}\mathcal{X}_{2} \label{Eq:3Node_Chain_Set_Constraints}
\end{align}
for the two generating elements $\mathcal{X}_{1}$ and $\mathcal{X}_{2}$. Using the constraints \eqref{Eq:3VertexChain_MinimalCondition1} implies
\begin{align}
    & g_{1,1}=\gcd\left(\ell_{1},k_{1}k_{2}\right)=\gcd\left(\ell_{1},k_{2}\right)=\gcd\left(\ell_{1}\ell_{2},k_{2}\right)=g_{2,2} \; .
\end{align}
In that case, the 3d $\Ncal=4$ mirror theory (see Figure~\ref{Fig:3VertexChain_Mirror}) is equipped with the discrete gauge group $\Z_{\gcd\left(\ell_{1},k_{2}\right)}$. There are different choices of fixing the charge assignment of the hypermultiplets $\{(\mathcal{A}_{i},\widetilde{\mathcal{A}_{i}})\}_{i=1}^{2}$ in the mirror dual theory that give equivalent representations / charge assignments. Here, the following set of charges, consistent with the  constraints \eqref{Eq:3Node_Chain_Set_Constraints} is chosen 
\begin{subequations}
\label{Eq:AbelianChainExample3Nodes_Charges}
    \begin{align}
    k_{1}\mathcal{X}_{1}&=\ell_{2}\mathcal{X}_{2} \quad \Leftrightarrow \quad \left[\mathcal{A}_{1}\right]=L_2 \quad \land \quad \left[\mathcal{A}_{2}\right]=K_1 \; , \\
 \text{with}\qquad   K_{1}&\equiv \frac{k_{1}}{g_{1}} \; , \quad L_{2}\equiv \frac{\ell_{2}}{g_{1}} \; , \quad g_{1}\equiv \gcd\left(\ell_{2},k_{1}\right) \; .
\end{align}
\end{subequations}
This gives the mirror dual theory\footnote{Another possible choice of charge assignment is to replace $L_{2}$, $K_{1}$ with $\ell_{2}$, $k_{1}$, respectively. For that, note that for a discrete theory of order $p$ with charge $q$ the following holds: $\Z^{[q]}_{p}\cong\Z^{[q]}_{p/\gcd\left(p,q\right)}$.} in Figure \ref{Fig:3VertexChain_Mirror}. 

Indeed, higgsing either of the (monomial) gauge-invariant operators formed from $\{(\mathcal{A}_{i},\widetilde{\mathcal{A}_{i}})\}_{i=1}^{2}$ leaves no residual gauge group. To see that, note that the residual gauge group associated with the higgsing of the hypermultiplet $(\mathcal{A}_{i},\widetilde{\mathcal{A}_{i}})$ is defined to be the biggest sub-group of $\Z_{\gcd\left(\ell_{1},k_{2}\right)}$ that leaves the hypermultiplet invariant, i.e. there is a non-trivial residual gauge group if and only if the charge of the hypermultiplet and the order-factor $\gcd\left(\ell_{1},k_{2}\right)$ share a common divisor. However, here\footnote{Recall the commutativity and associativity property of the $\gcd$-function.}:
\begin{subequations}
\begin{align}
    & \gcd\left(\ell_{2},\gcd\left(\ell_{1},k_{2}\right)\right)=\gcd\big(\underbrace{\gcd\left(\ell_{2},k_{2}\right)}_{\stackrel{\eqref{Eq:3VertexChain_MinimalCondition1}}{=}1},\ell_{1}\big)=1 \; ,
    \\
    & \gcd\left(k_{1},\gcd\left(\ell_{1},k_{2}\right)\right)=\gcd\big(\underbrace{\gcd\left(k_{1},\ell_{1}\right)}_{\stackrel{\eqref{Eq:3VertexChain_MinimalCondition1}}{=}1},k_{2}\big)=1 \; ,
\end{align}
\end{subequations}
that is, there is no non-trivial higgsing possible. Thus, the mirror theory exhibits a minimal Higgs branch, and, by extension, the chain quiver a minimal Coulomb branch. The geometry of this minimal Coulomb branch is identified with
\begin{align}
    \mathcal{M}_{\Coulomb} \cong \left( \mathbb{H}^{2}/\Z_{\gcd\left(\ell_{1},k_{2}\right)} \right)^{\left[K_{1},L_{2}\right]}
   \stackrel{\eqref{eq:h-def}}{\equiv}
   h_{2,\gcd\left(\ell_{1},k_{2}\right),(K_{1},L_{2})}
    \;,
\end{align}
with $\left[K_{1},L_{2}\right]$ the charges with respect to the discrete group action.

To sum up, the following holds for the three-vertices chain quiver (Figure~\ref{Fig:3VertexChain_Quiver}): 
\begin{align}
      \left( \substack{\text{three-vertices chain}  \\ \text{(Figure~\ref{Fig:3VertexChain_Quiver})}  }\right) =
    \begin{cases}
        \substack{ \text{non-trivial theory} \\ \text{and minimal Coulomb branch $\mathcal{M}_{\Coulomb}$}} & \Leftrightarrow  \gcd\left(\ell_{i},k_{i}\right)=1 \land \gcd\left(\ell_{1},k_{2}\right)>1 \; , \\ 
        \\
        \substack{\text{trivial theory} \\  \text{(and trivial Coulomb branch $\mathcal{M}_{\Coulomb}$)}} & \Leftrightarrow  \gcd\left(\ell_{i},k_{i}\right)=1 \land \gcd\left(\ell_{1},k_{2}\right)=1 \; .
    \end{cases}
\end{align}

\paragraph{Abelian chain quivers with $n$-many vertices.} In order to close the induction, assume now that the chain quiver with $(n-1)$-many vertices and $\left(\ell,-k\right)$-edges as in Figure \ref{Fig:Complete_Class_AbelianChain_Minimal} is minimal if and only if the constraints \eqref{Eq:AbelianChain_GCD_Condition} are fulfilled, i.e.
\begin{subequations}
\begin{alignat}{2}
    & \gcd\left(\ell_{1},k_{j}\right)=1 & \quad &\forall \, j: \, 1\leq j\leq n-3 \; , \label{Eq:NAbelianChain_GCD_Condition1}
    \\
    & \gcd\left(\ell_{i},k_{n-1}\right)=1 & \quad &\forall \, i: \, 2\leq i\leq n-2 \; , \label{Eq:NAbelianChain_GCD_Condition2}
    \\
    & \gcd\left(\ell_{i},k_{j}\right)=1 & \quad &\forall \, i,j: \, 2\leq i\leq j\leq n-3 \; , \label{Eq:NAbelianChain_GCD_Condition3}
    \\
    & \gcd\left(\ell_{1},k_{n-2}\right)>1 \;,  & & \label{Eq:NAbelianChain_GCD_Condition4}
\end{alignat} \label{Eq:NAbelianChain_GCD_Conditions}
\end{subequations}
and that the chain is trivial if and only if the constraint \eqref{Eq:NAbelianChain_GCD_Condition4} is trivialized, i.e.
\begin{align}
    \gcd\left(\ell_{1},k_{n-2}\right)=1 \; . \label{Eq:NAbelianChain_TrivialGCD}
\end{align}
Consider now the chain quiver with $n$-many vertices and $\left(\ell,-k\right)$-edges as in Figure \ref{Fig:Complete_Class_AbelianChain_Minimal} and 
\begin{sidewaysfigure}
    \centering
    \includegraphics[page=1]{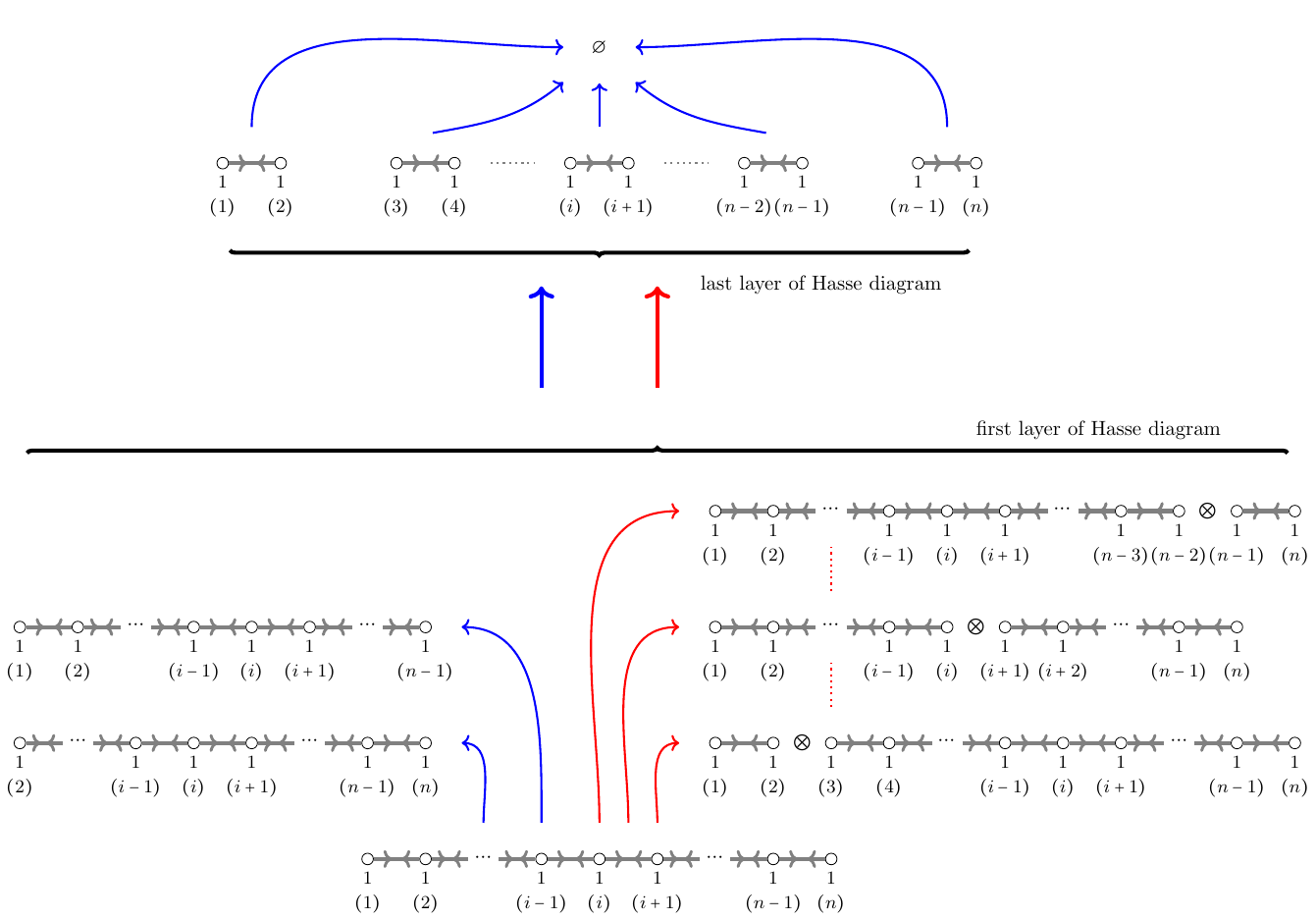}
\caption{The first and last layer of the generic Hasse diagram for the generic $n$-vertices chain quiver. 
The red-colored and blue-colored arrows denote fission and decay-channels; of which the $n$-vertices chain has $(n-3)$ and $2$-many, respectively, at the first layer of the stratification. Note that in the following layers the different channels can intertwine and that the last layer consists of all possible $2$-vertices sub-chains that are embedded in the initial $n$-vertices chain.}
\label{Fig:NVertexChain_Hasse}
\end{sidewaysfigure}
its (generic) Hasse diagram sketched in Figure~\ref{Fig:NVertexChain_Hasse}. Here, focus on the first layer of decay and fission products. The $n$-vertices chain has two decay channels resulting in two $\left(n-1\right)$-vertices chains, and $\left(n-3\right)$-many fission channels resulting in products of $p$-vertices chains with $\left(n-p\right)$-vertices chains $\left(2\leq p\leq n-2\right)$. Note that these products consist of chain quivers that appear in the Hasse diagram of the two $\left(n-1\right)$-vertices decay products. That is to say, with the assumption that the two $\left(n-1\right)$-vertices chains are trivial, these $p$-vertices chains are necessarily trivial as well. Therefore, the Hasse diagram for the initial $n$-vertices chain trivializes.

\begin{figure}[t!]
\centering
\includegraphics[page=1]{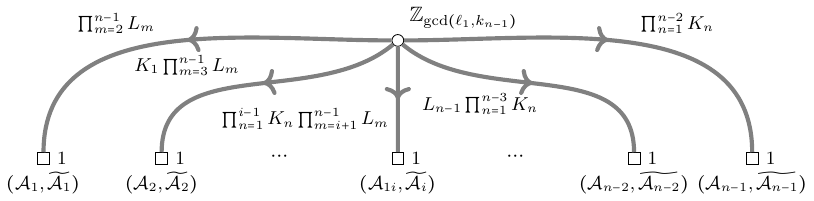}
\caption{The mirror dual theory to the $n$-vertices chain (Figure \ref{Fig:Complete_Class_AbelianChain_Minimal}), provided that both embedded $(n-1)$-vertices chains satisfy the constraints \eqref{Eq:NAbelianChain_GCD_Condition1}, \eqref{Eq:NAbelianChain_GCD_Condition2}, \eqref{Eq:NAbelianChain_GCD_Condition3}, and \eqref{Eq:NAbelianChain_TrivialGCD}. The mirror dual has a discrete gauge group $\Z_{\gcd\left(\ell_{1},k_{n-1}\right)}$ and $\left(n-1\right)$-many $\Ncal=2$ chiral-pairs $(\mathcal{A}_{i},\widetilde{\mathcal{A}_{i}})$ with varying gauge charges; see \eqref{Eq:NAbelianChain_Charges_all}.}
\label{Fig:NVertexChain_Mirror}
\end{figure}

In more formal terms, consider the 3d $\Ncal=4$ mirror theory to the $n$-vertices chain. For that, recall the charge matrix \eqref{Eq:NChain_Charge} and the relations \eqref{Eq:General_AbelianChain_Relations_Charge} and \eqref{Eq:General_AbelianChain_Relations_GCD} defining the cokernel spanned by the generating set $\left\{\mathcal{X}_{i}\right\}_{i=1}^{n-1}$. Enforcing the constraints \eqref{Eq:NAbelianChain_GCD_Condition1}, \eqref{Eq:NAbelianChain_GCD_Condition2}, \eqref{Eq:NAbelianChain_GCD_Condition3} and \eqref{Eq:NAbelianChain_TrivialGCD} for the two $\left(n-1\right)$-vertices chains embedded in the $n$-vertices chain, i.e.\ the sub-chain starting on node $\left(1\right)$ and ending on node $\left(n-1\right)$ and the sub-chain starting on node $\left(2\right)$ and ending on node $\left(n\right)$ --- which is exactly enforcing the constraints \eqref{Eq:AbelianChain_GCD_Condition1}-\eqref{Eq:AbelianChain_GCD_Condition3} --- yields for \eqref{Eq:GeneralAbelianChain_GCD_Function}
\begin{align}
    g_{p,q}=\gcd\left(\prod_{j=1}^{p}\ell_{j},\prod_{m=q}^{n-1}k_{m}\right)=\gcd\left(\ell_{1},k_{n-1}\right)=g_{1,n-1} \quad \left(\forall \,p,q: \, 1\leq p,q \leq n-1\right) \; .
\end{align}
Thus, the relation \eqref{Eq:General_AbelianChain_Relations_GCD} implies the gauge group of the mirror theory to be $\Z_{\gcd\left(\ell_{1},k_{n-1}\right)}$. As before, there are different choices of fixing the charges that define equivalent representations / charge assignments of the same mirror dual theory. Here, the following charge assignment consistent with the set of relations \eqref{Eq:General_AbelianChain_Relations_Charge} is chosen:
\begin{subequations}
\label{Eq:NAbelianChain_Charges_all}
\begin{alignat}{2}
    \left[\mathcal{A}_{1}\right]&=\prod_{m=2}^{n-1}L_{m} \; , \quad \left[\mathcal{A}_{n-1}\right]&=\prod_{n=1}^{n-2}K_{n} \; ,
    \\
    \left[\mathcal{A}_{i}\right]&=\prod_{m=i+1}^{n-1}L_{m}\prod_{n=1}^{i-1}K_{n} \label{Eq:NAbelianChain_Charges} & \quad &\forall \, i: \, 2\leq i\leq n-2 \\
  \text{with} \qquad   K_{i}&\equiv k_{i}/g_{i} & \quad &\forall \, i: \, 1\leq i \leq n-2 \; ,
    \\
     L_{j}&\equiv \ell_{j}/g_{j-1}& \quad &\forall \, j: \, 2\leq j \leq n-1 \; ,
    \\
    g_{p}&\equiv \gcd\left(\ell_{p+1},k_{p}\right) &\quad &\forall \, p: \, 1\leq p \leq n-2 \; .
\end{alignat}
\end{subequations}
Put together, this yields the 3d $\Ncal=4$ mirror theory\footnote{Note, as in the previous case, another choice of gauge charges by replacing $L_{i}$, $K_{i}$ with $\ell_{i}$, $k_{i}$, respectively.} shown in Figure \ref{Fig:NVertexChain_Mirror}.

The mirror dual theory indeed does not allow for a non-trivial higgsing. To see that, consider higgsing any of the (monomial) gauge-invariant operators that can be constructed using the set of hypermultiplets $\{(\mathcal{A}_{i},\widetilde{\mathcal{A}_{i}})\}_{i=1}^{n-1}$, for which one has
\begin{subequations}
\begin{align}
    \gcd\left(\prod_{m=2}^{n-1}\ell_{m},\gcd\left(\ell_{1},k_{n-1}\right)\right)&\stackrel{\phantom{\eqref{Eq:AbelianChain_GCD_Condition1}}}{=}\gcd\Bigg(\underbrace{\gcd\left(\prod_{m=2}^{n-1}\ell_{m},k_{n-1}\right)}_{\stackrel{\eqref{Eq:AbelianChain_GCD_Condition2}}{=1}},\ell_{1}\Bigg)=1
    \\
    \gcd\left(\prod_{n=1}^{n-2}k_{n},\gcd\left(\ell_{1},k_{n-1}\right)\right)&\stackrel{\phantom{\eqref{Eq:AbelianChain_GCD_Condition1}}}{=}\gcd\Bigg(\underbrace{\gcd\left(\prod_{n=1}^{n-2}k_{n},\ell_{1}\right)}_{\stackrel{\eqref{Eq:AbelianChain_GCD_Condition1}}{=1}},k_{n-1}\Bigg)=1 
    \\
    \gcd\left(\prod_{m=i+1}^{n-1}\ell_{m}\prod_{n=1}^{i-1}k_{n},\gcd\left(\ell_{1},k_{n-1}\right)\right)&\stackrel{\phantom{\eqref{Eq:AbelianChain_GCD_Condition1}}}{=}\gcd\left(\gcd\left(\prod_{m=i+1}^{n-1}\ell_{m}\prod_{n=1}^{i-1}k_{n},\ell_{1}\right),k_{n-1}\right) 
    \\
    &\stackrel{\eqref{Eq:AbelianChain_GCD_Condition1}}{=}\gcd\left(\gcd\left(\prod_{m=i+1}^{n-1}\ell_{m},\ell_{1}\right),k_{n-1}\right) \notag
    \\
    &\stackrel{\phantom{\eqref{Eq:AbelianChain_GCD_Condition1}}}{=}\gcd\Bigg(\ell_{1},\underbrace{\gcd\left(\prod_{m=i+1}^{n-1}\ell_{m},k_{n-1}\right)}_{\stackrel{\eqref{Eq:AbelianChain_GCD_Condition2}}{=1}}\Bigg)=1 \quad\forall \, i: \, 2\leq i\leq n-2 \; . \notag
\end{align}    
\end{subequations}

In conclusion, the $n$-vertices chain of Figure \ref{Fig:Complete_Class_AbelianChain_Minimal} is non-trivial and minimal if and only if
\begin{subequations}
\begin{alignat}{2}
    & \gcd\left(\ell_{1},k_{j}\right)=1 & \quad &\forall \, j: \, 1\leq j\leq n-2 \; ,
    \\
    & \gcd\left(\ell_{i},k_{n-1}\right)=1 & \quad &\forall \, i: \, 2\leq i\leq n-1 \; ,
    \\
    & \gcd\left(\ell_{i},k_{j}\right)=1 & \quad &\forall \, i,j: \, 2\leq i\leq j\leq n-2 \; ,
    \\
    & \gcd\left(\ell_{1},k_{n-1}\right)>1 \; , & &
\end{alignat}
\end{subequations}
and has a Coulomb branch defined by the orbifold
\begin{align}
    \mathcal{M}_{\Coulomb} \cong \left( \mathbb{H}^{n-1}/\Z_{\gcd\left(\ell_{1},k_{n-1}\right)} \right)^{[\sigma]}
    \stackrel{\eqref{eq:h-def}}{\equiv}
    h_{n-1,\gcd\left(\ell_{1},k_{n-1}\right),\sigma}
    \label{Eq:CB_min_chain}
\end{align}
with the vector of charges $\sigma$ as defined in \eqref{Eq:NAbelianChain_Charges}. The interpretation of \eqref{Eq:CB_min_chain} as Higgs branch of the mirror theory provides explicit confirmation that the established minimal $n$-vertices chain is a good Abelian quiver theory.

\subsubsection{Stable Abelian chain quivers in reduced representation}
\label{Sec:Abelian_Chain_Equivalence}

\begin{figure}[t!]
\begin{subfigure}[t]{1\textwidth}
    \centering
    \includegraphics[page=1]{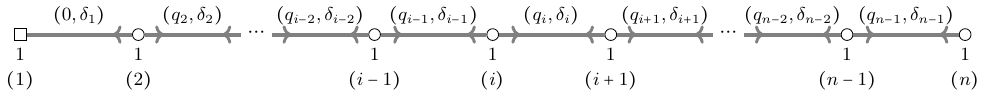}
\caption{}
\label{Fig:NVertexChain_Quiver_Reduced}
\end{subfigure}
\begin{subfigure}[t]{1\textwidth}
    \centering
    \includegraphics[page=2]{Abelian_Decay_Fission_Fig12.pdf}
\caption{}
\label{Fig:NVertexChain_Quiver_Reduced_Minimal}
\end{subfigure}
\caption{Figure \subref{Fig:NVertexChain_Quiver_Reduced} showcases the reduced representation of the Abelian $n$-vertices chain with $\left(\ell,-k\right)$-edges in Figure \ref{Fig:Complete_Class_AbelianChain_Minimal}. Note the relations between both representations $\delta_i\equiv\gcd\left(k_{i},\prod_{j=1}^{i}\ell_{j}\right)=u_{i}\left(\prod_{j=1}^{i}\ell_{j}\right)-\upsilon_{i}k_{i}$ $\left(1\leq i \leq n-1\right)$ and $q_{j}\equiv\ell_{j}\upsilon_{j-1}$ $\left(2\leq j\leq n-1\right)$, with $u_{i}$, $\upsilon_{i}$ a choice of Bézout coefficients of the $\gcd$-expression. By construction, $q_{1}=0$, which yields the framing on vertex with label $\left(1\right)$. Figure \subref{Fig:NVertexChain_Quiver_Reduced_Minimal} showcases the same quiver in the configuration for which the Coulomb branch exhibits a trivial stratification.}
\label{Fig:NVertexChain_Reduced}
\end{figure}

Recall from Section~\ref{Sec:Quiver_reduced_form} the reduced quiver representation (Figure~\ref{Fig:NVertexChain_Quiver_Reduced}) for the $n$-vertices chain with charges:
\begin{subequations}
\label{Eq:Bezout_coeffs}
\begin{alignat}{2}
    & \delta_{i}\equiv\gcd\left(k_{i},\prod_{j=1}^{i}\ell_{j}\right)=u_{i}\left(\prod_{j=1}^{i}\ell_{j}\right)-\upsilon_{i}k_{i} & \qquad &\forall\,i:\, 1\leq i \leq n-1 \; , \label{Eq:Bezout_coeffs_1}
    \\
    & q_{j}\equiv \ell_{j}\upsilon_{j-1} & \qquad &\forall\,j:\, 2\leq j\leq n-1 \; ,\label{Eq:Bezout_coeffs_2}
\end{alignat}
\end{subequations}
with $u_{i}$ and $\upsilon_{i}$ a choice of Bézout coefficients for the decomposition of the $\gcd$-expression. Recall that by construction $q_{1}=0$ in the reduced representation, yielding the framing on vertex with label $\left(1\right)$. 

\paragraph{Cokernel for reduced quiver.}
Before starting the proof, first determine the (schematic) 3d $\Ncal=4$ mirror theory of the reduced form in Figure~\ref{Fig:NVertexChain_Quiver_Reduced}. The charge matrix $\rho: \Z^{n-1}\rightarrow\Z^{n-1}$ reads  
\begin{equation}
\rho=
\begin{tikzpicture}[baseline]
\matrix (m) [matrix of math nodes,nodes in empty cells,right delimiter={)},left delimiter={(} ]{
\delta_1 & 0 & 0 & & & & & 0 \\
q_2 & \delta_2 & 0 & & & & & 0 \\
0 & q_3 & \delta_3 & & & & & 0 \\
& & & & & & &  \\
& & & & & & &  \\
& & & & & & &  \\
0 & & & & & q_{n-2} & \delta_{n-2} & 0\\
0 & & & & & 0 & q_{n-1} & \delta_{n-1}\\
} ;
\draw[loosely dotted, thick] (m-1-3)-- (m-1-8);
\draw[loosely dotted, thick] (m-3-8)-- (m-7-8);
\draw[loosely dotted, thick] (m-3-1)-- (m-7-1);
\draw[loosely dotted, thick] (m-8-1)-- (m-8-6);
\draw[loosely dotted, thick] (m-3-3)-- (m-7-7);
\end{tikzpicture} \; . \label{Eq:NChain_Charge_Reduced}
\end{equation}
The generating set $\left\{\mathcal{X}_{i}\right\}_{i=1}^{n-1}$ of $\Z^{n-1}/\mathrm{Im}\left(\rho\right)$ is defined via the relations
\begin{subequations}
\label{Eq:NAbelianChain_Reduced_Cokernel1}
\begin{align}
     \delta_{n-1}\mathcal{X}_{n-1}&=0 \; , \label{Eq:NAbelianChain_Reduced_Cokernel_n-1}
    \\
     \delta_{i}\mathcal{X}_{i}&=-q_{i+1}\mathcal{X}_{i+1} \qquad \forall \, i: \, 1\leq q\leq n-2 \; .
\end{align}
\end{subequations}
Put together, one finds
\begin{align}
    \prod_{j=i}^{n-1}\delta_{j}\mathcal{X}_{i}=0 \qquad \forall \, i: \, 1\leq i \leq n-1 \; . \label{Eq:NAbelianChain_Reduced_Cokernel2}
\end{align}

\begin{figure}[t!]
    \centering
    \includegraphics[page=1]{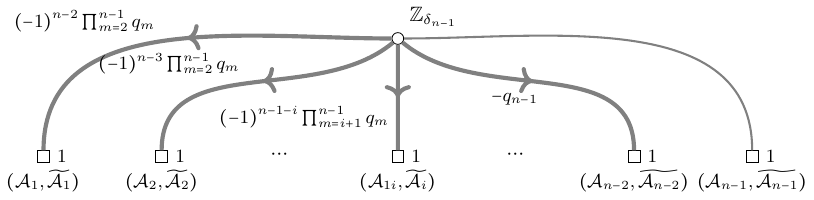}
\caption{The 3d $\Ncal=4$ mirror theory to the $n$-vertices chain (Figure~\ref{Fig:NVertexChain_Quiver_Reduced}), provided \eqref{Eq:NAbelianChain_Reduced_Minimal} holds.}
\label{Fig:NVertexChain_Quiver_Reduced_Mirror}
\end{figure}

\paragraph{From gcd conditions to minimal Coulomb branch.} Now, the aim is to show that imposing the gcd-conditions \eqref{Eq:General_AbelianChain_Relations_GCD} on the reduced chain implies the mirror to have a minimal Higgs branch. In this reduced representation, the constraints simplify to
\begin{subequations}
\label{Eq:NAbelianChain_Reduced_GCD}
\begin{align}
    & \delta_{i}=\gcd\left(k_{i},\prod_{j=1}^{i}\ell_{j}\right)=1 \qquad \forall \, i: 1\leq i\leq n-2 \; , \label{Eq:NAbelianChain_Reduced_GCD1}
    \\
    & \delta_{n-1}=\gcd\left(k_{n-1},\ell_{1}\right)>1 \; . \label{Eq:NAbelianChain_Reduced_GCD2}
\end{align}
\end{subequations}
Assuming the constraints \eqref{Eq:NAbelianChain_Reduced_GCD} to hold, the chain takes the form of Figure~\ref{Fig:NVertexChain_Quiver_Reduced_Minimal} and, similar to the proof via induction, the relation \eqref{Eq:NAbelianChain_Reduced_Cokernel2} simplifies to
\begin{subequations}
\label{Eq:NAbelianChain_Reduced_Minimal}
\begin{alignat}{2}
     \delta_{n-1}\mathcal{X}_{i}&=0 &\qquad  &\forall \, i: 1\leq i \leq n-1\; ,
    \\
    \text{and furthermore} \qquad 
     \mathcal{X}_{i}&=(-1)^{n-1-i}\prod_{j=+1}^{n-1}q_{j}\mathcal{X}_{n-1} &\qquad  &\forall \, i: 1\leq i \leq n-2 \; .
\end{alignat}
\end{subequations}
In conclusion, the 3d $\Ncal=4$ mirror theory is realized by the orbifold theory in Figure~\ref{Fig:NVertexChain_Quiver_Reduced_Mirror},
where the charge assignment is fixed by setting the charges of the $\Ncal=2$ chiral pair $(\mathcal{A}_{n-1},\widetilde{\mathcal{A}_{n-1}})$ to $1$. In the Appendix \ref{Sec:Appendix_Mirror_Matching} it is explicitly shown that this representation of the mirror theory for the $n$-vertices chain with minimal Coulomb branch is equivalent to the one presented in Section \ref{Sec:Abelian_Chain_Induction}.
 
\paragraph{From minimal Coulomb branch to gcd conditions.}
For the proof in the converse direction, assume that one is given an $n$-vertices chain, as in Figure~\ref{Fig:NVertexChain_Reduced}, which has a minimal Coulomb branch.

Assume, contrary to the constraint \eqref{Eq:NAbelianChain_Reduced_GCD2}, $ \delta_{n-1}=1$.
However, the relation \eqref{Eq:NAbelianChain_Reduced_Cokernel_n-1} then trivializes to $ \mathcal{X}_{n-1}=0$.
Thus, the associated hypermultiplet decouples from the theory and the Higgs branch of the mirror theory (and hence the Coulomb branch of the chain) cannot be minimal. In order to remedy this contradiction, one indeed needs $ \delta_{n-1}>1$.

Consider now, contrary to the constraints \eqref{Eq:NAbelianChain_Reduced_GCD1}, that there is exactly one $i$ for $1\leq i\leq n-2$ such that $\delta_{i}>1$ (one can straightforwardly generalize to at least one $i$ such that $\delta_i>1$). From the relations \eqref{Eq:NAbelianChain_Reduced_Cokernel1}, \eqref{Eq:NAbelianChain_Reduced_Cokernel2}, and \eqref{Eq:NAbelianChain_Reduced_Cokernel_n-1} one finds
\begin{subequations}
\label{Eq:NAbelianChain_Reduced_Constraints}
\begin{alignat}{2}
     \delta_{i}\delta_{n-1}\mathcal{X}_{j}&=0 \qquad & &\forall \, j: \, 1\leq j\leq i \; , \label{Eq:NAbelianChain_Reduced_AlternativeProof_Degree1}
    \\
     \delta_{n-1}\mathcal{X}_{j}&=0 \qquad & &\forall \, j: \, i< j \leq n-1 \; , \label{Eq:NAbelianChain_Reduced_AlternativeProof_Degree2} \\ 
     \delta_{i}\mathcal{X}_{j}&=\prod_{k=j+1}^{n-1}q_{k}\mathcal{X}_{n-1} \qquad & &\forall \, j: \, 1\leq j\leq i \; , \label{Eq:NAbelianChain_Reduced_AlternativeProof_Charge2}
    \\
     \mathcal{X}_{j}&=\prod_{k=j+1}^{n-1}q_{k}\mathcal{X}_{n-1} \qquad & &\forall \, j: \, i< j<n-1 \; . \label{Eq:NAbelianChain_Reduced_AlternativeProof_Charge1}
     \\
     \mathcal{X}_{j}&=\prod_{k=j+1}^{i}q_{k}\mathcal{X}_{i} \qquad & &\forall\, j:\, 1\leq j\leq i \;. \label{Eq:NAbelianChain_Reduced_AlternativeProof_Charge3}
\end{alignat}
\end{subequations}
Generically, using the constraints \eqref{Eq:NAbelianChain_Reduced_AlternativeProof_Degree1} and \eqref{Eq:NAbelianChain_Reduced_AlternativeProof_Degree2}, the gauge group $G$ is a product of discrete factors in the form $G=\prod_{p_{i}}\Z_{p_{i}}$. Consider the following two distinct cases: Assume that the gauge group consists of only one factor\footnote{With the lower bound $p\geq\delta_{i}\delta_{n-1}$.}, i.e.\ $G=\Z_{p}$ with $\Z_{\delta_{n-1}}\subset\Z_{\delta_{i}\delta_{n-1}}\subset\Z_{p}$ due to the constraints \eqref{Eq:NAbelianChain_Reduced_AlternativeProof_Degree1} and \eqref{Eq:NAbelianChain_Reduced_AlternativeProof_Degree2}. In that case, higgsing the hypermultiplet associated with the generating element $\mathcal{X}_{n-1}$ leaves at the very least the residual gauge group $\Z_{\delta_{n-1}}$, thus, violating the assumption that the Coulomb branch is minimal.

\begin{figure}[t!]
    \centering
    \includegraphics[page=1]{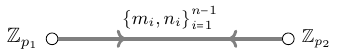}
\caption{3d $\Ncal=4$ quiver theory with a product gauge group consisting of two distinct factors. The label $\left\{m_{i},n_{i}\right\}_{i=1}^{n-1}$ denotes $\left(n-1\right)$-many $\Ncal=4$ hypermultiplets with integer-valued gauge charges $m_{i}$ and $n_{i}$ with respect to the gauge factors $\Z_{p_{1}}$ and $\Z_{p_{2}}$, respectively. For this quiver theory to exhibit a minimal Coulomb branch, the $gcd$ of the charges with their respective order-factor $p_{1}$ or $p_{2}$ needs to be trivial.}
\label{Fig:NAbelianChain_ReducedForm_ProductCase}
\end{figure}

Instead of a single factor, assume now that the gauge group is a product group of at least two factors. For the case at hand, such a factorization of the gauge group can only be encoded in the constraint \eqref{Eq:NAbelianChain_Reduced_AlternativeProof_Charge2}, separating the two ``sectors'' encoded in \eqref{Eq:NAbelianChain_Reduced_AlternativeProof_Charge1} and \eqref{Eq:NAbelianChain_Reduced_AlternativeProof_Charge3}. Therefore, the gauge group product can only consist of two factors\footnote{With the lower bounds $p_{1}\geq\delta_{n-1}$ and $p_{2}\geq\delta_{i}\delta_{n-1}$}:
\begin{align}
    G&=\Z_{p_{1}}\times\Z_{p_{2}}
\qquad 
\text{with}
\qquad
    \Z_{\delta_{n-1}}\subset\Z_{p_{1}} \quad \text{and} \quad \Z_{\delta_{i}\delta_{n-1}}\subset\Z_{p_{2}} \; .
\end{align}
For a connected quiver with such a gauge group and a minimal Coulomb branch, the quiver has to take the form\footnote{Note that the assumption here is that the mirror dual in the case of a stable theory can be expressed as a quiver theory. However, one can also make the argument without this assumption.} of Figure \ref{Fig:NAbelianChain_ReducedForm_ProductCase}
with the condition
\begin{align}
    \gcd\left(m_{i},p_{1}\right)=1 \quad \land \quad \gcd\left(n_{i},p_{2}\right)=1 \qquad \forall \, i: \, 1\leq i\leq n-1 \;.
\end{align}
This implies
\begin{align}
    \gcd\left(m_{i},\delta_{n-1}\right)=1 \quad \land \quad \gcd\left(n_{i},\delta_{i}\delta_{n-1}\right)=1 \qquad \forall \, i: \, 1\leq i\leq n-1 \; .
\end{align}
However, in \eqref{Eq:NAbelianChain_Reduced_AlternativeProof_Charge3} included is the relation $\mathcal{X}_{i-1}=q_{i}\mathcal{X}_{i}$ with $q_{i}=\ell_{i}\upsilon_{i-1} $
such that the charge of the $\Ncal=2$ chiral pair $(\mathcal{A}_{i-1},\widetilde{\mathcal{A}_{i-1}})$ associated with the generating element $\mathcal{X}_{i-1}$ reads $ \left[\mathcal{A}_{i-1}\right]=q_{i}\left[\mathcal{A}_{i}\right] $. For this charge\footnote{Note the slight abuse of notation: The charge $\left[\mathcal{A}_{i}\right]$ includes a charge for both gauge factors. Therefore, in the $\gcd$-functions, it would be enough to consider just the relative factor $q_{i}$.} one finds
\begin{align}
    \gcd\left(\left[\mathcal{A}_{i-1}\right],\delta_{i}\right)&=\gcd\left(\ell_{i}\upsilon_{i-1}\left[\mathcal{A}_{i}\right],\gcd\left(\ell_{i},\prod_{j=1}^{i}k_{j}\right)\right)
    \\
    &=\gcd\left(\gcd\left(\ell_{i}\upsilon_{i-1}\left[\mathcal{A}_{i}\right],\ell_{i}\right),\prod_{j=1}^{i}k_{j}\right) \nonumber
    \\
    &=\gcd\left(\ell_{i},\prod_{j=1}^{i}k_{j}\right)=\delta_{i}>1 \; , \nonumber
    \\
    \gcd\left(\left[\mathcal{A}_{i-1}\right],\delta_{i}\delta_{n-1}\right)&\geq\delta_{i} \; .
\end{align}
Therefore, higgsing the hypermultiplet associated with the generating element $\mathcal{X}_{i-1}$ leaves a non-trivial gauge group, violating the assumption that the quiver theory has a minimal Coulomb branch.

This argument can be extended to $\delta_{i}>1$ $\forall \, i: \, 1\leq i\leq n-2$. In conclusion, in order for the Coulomb branch of the Abelian $n$-vertex chain to be minimal, one indeed requires
\begin{align}
    \delta_{i}=1 \quad \forall \, i: \, 1\leq i\leq n-2 \; .
\end{align}

\subsection{Tree-like quivers beyond the chain}
\label{Sec:AbelianTreeExtension}
\begin{figure}[t!]
    \centering
    \includegraphics[page=1]{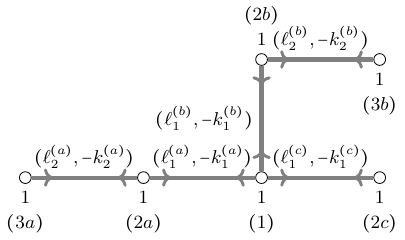}
\caption{Example of a tree-like extension of an Abelian chain quiver theory.}
\label{Fig:Tree_Extension_Example}
\end{figure}

In this section it is proven that there exist no Abelian tree-like quiver theories beyond the chains (Figure \ref{Fig:Complete_Class_AbelianChain_Minimal}) that are non-trivial and minimal simultaneously. 
To illustrate the argument, consider first the Abelian quiver of Figure~\ref{Fig:Tree_Extension_Example}, which is defined via the charge matrix $  \rho: \Z^{6}\mapsto \Z^{5}$ with
\begin{align}
    \rho=   \begin{pmatrix}
            \ell^{(a)}_{2} & -k^{(a)}_{2} & 0 & 0 & 0 & 0\\
            0 & \ell^{(a)}_{1} & -k^{(a)}_{1} & 0 & 0 & 0\\
            0 & 0 & \ell_{1}^{(b)} & -k_{1}^{(b)} & 0 & 0\\
            0 & 0 & 0 & \ell_{(2)}^{(b)} & -k_{2}^{(b)} & 0\\
            0 & 0 & \ell_{1}^{(c)} & 0 & 0 & -k_{1}^{(c)} \\
            \end{pmatrix} \; .
\end{align}
The cokernel (and hence the 3d $\Ncal=4$ mirror theory) yields the following constraints for the generating set $\left\{\mathcal{X}^{(a)}_{1}, \mathcal{X}^{(a)}_{2}, \mathcal{X}^{(b)}_{1}, \mathcal{X}^{(b)}_{2}, \mathcal{X}^{(c)}_{1}\right\}$
\begin{subequations}
\begin{alignat}{3}
     \ell^{(a)}_{2}\mathcal{X}^{(a)}_{2}&=0 \; , \label{Eq:Tree_Extension_ExampleFree}
    & \qquad 
     k_{2}^{(b)}\mathcal{X}^{(b)}_{2}&=0 \; , 
    & \qquad
     k_{1}^{(c)}\mathcal{X}^{(c)}_{1}&=0 \; , 
    \\
     k^{(a)}_{2}\mathcal{X}^{(a)}_{2}&=\ell^{(a)}_{1}\mathcal{X}^{(a)}_{1} \; ,
    &\qquad k^{(a)}_{1}\mathcal{X}^{(a)}_{1}&=\ell_{1}^{(b)}\mathcal{X}^{(b)}_{1}+\ell_{1}^{(c)}\mathcal{X}^{(c)}_{1} \; ,
    &\qquad
    k_{1}^{(b)}\mathcal{X}^{(b)}_{1}&=\ell_{2}^{(b)}\mathcal{X}^{(b)}_{2} \; .
\end{alignat}
\end{subequations}
Combining these, one finds for the generating elements $\mathcal{X}^{(a)}_{2}$, $\mathcal{X}^{(b)}_{2}$ and $\mathcal{X}^{(c)}_{1}$ the additional constraints
\begin{align}
\label{Eq:Tree_Extension_ExampleFree4}
    & k_{2}^{(b)}k_{1}^{(b)}k_{1}^{(c)}k_{1}^{(a)}k_{2}^{(a)}\mathcal{X}_{2}^{(a)}=0 \; , 
   \qquad 
   \ell_{2}^{(a)}\ell_{1}^{(a)}k_{1}^{(c)}\ell_{1}^{(c)}\ell_{2}^{(b)}\mathcal{X}_{2}^{(b)}=0 \; ,
    \qquad
     k_{2}^{(b)}k_{1}^{(b)}\ell_{2}^{(a)}\ell_{1}^{(a)}\ell_{1}^{(c)}\mathcal{X}_{1}^{(c)}=0 \; .
\end{align}
Putting \eqref{Eq:Tree_Extension_ExampleFree} together with \eqref{Eq:Tree_Extension_ExampleFree4} yields for the three generating elements $\mathcal{X}^{(a)}_{2}$, $\mathcal{X}^{(b)}_{2}$ and $\mathcal{X}^{(c)}_{1}$ the following relations
\begin{subequations}
\label{Eq:Tree_Extension_Example_MaximalOrder}
\begin{align}
    & \gcd\left(\ell^{(a)}_{2},k_{1}^{(a)}k^{(a)}_{2}k_{1}^{(b)}k_{2}^{(b)}k^{(c)}_{1}\right)\mathcal{X}^{(a)}_{2}=0 \; ,
    \\
    & \gcd\left(k_{2}^{(b)},\ell_{1}^{(a)}\ell_{2}^{(a)}\ell_{2}^{(b)}\ell_{1}^{(c)}k_{1}^{(c)}\right)\mathcal{X}^{(b)}_{2}=0 \; ,
    \\
    & \gcd\left(k_{1}^{(c)},\ell_{1}^{(a)}\ell_{2}^{(a)}k_{1}^{(b)}k_{2}^{(b)}\ell_{1}^{(c)}\right)\mathcal{X}^{(c)}_{1}=0 \; .
\end{align}
\end{subequations}
Consider the form of the quiver shown in Figure~\ref{Fig:Tree_Extension_Example}: The decay and fission algorithm yields a generic Hasse diagram composed of all decays and fissions to all embedded sub-chain quivers and their products. In order to minimize this Hasse diagram, one needs to exclude these chains by forcing them to be trivial theories. As before, this is done by demanding the constraints \eqref{Eq:AbelianChain_GCD_Condition} to hold, with the modification that the constraint \eqref{Eq:AbelianChain_GCD_Condition4} trivializes. Doing so, yields the following trivialized $\gcd$-expressions
\begin{subequations}
\begin{alignat}{2}
     \gcd\left(\ell_{i}^{(m)},k_{i}^{(m)}\right)&=1 \qquad & &\forall \, i: \, 1\leq i\leq 2 \;, \; \forall \, m\in\left\{a,b,c\right\} \;,
    \\
    \gcd\left(\ell_{2}^{(a)},k_{i}^{(m)}\right)&=1 \qquad & &\forall \, i: \, 1\leq i\leq 2 \;, \; \forall \, m\in\left\{b,c\right\} \; ,
    \\
    \gcd\left(k_{1}^{(c)},k_{i}^{(b)}\right)&=1 \qquad & &\forall \, i: \, 1\leq i\leq 2 \;,
\end{alignat}
\end{subequations}
Thus, the $\gcd$-expressions in \eqref{Eq:Tree_Extension_Example_MaximalOrder} trivialize and the three hypermultiplets become free and decouple from the theory, i.e.\
\begin{align}
    \mathcal{X}_{1}=0, \quad \mathcal{X}_{4}=0, \quad \mathcal{X}_{5}=0 \; .
\end{align}

\begin{figure}[t!]
    \centering
    \includegraphics[page=2]{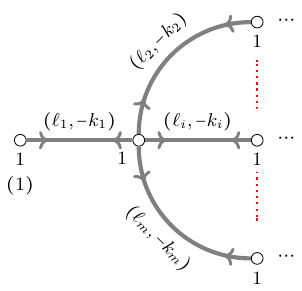}
\caption{Schematic section of a connected tree-like Abelian quiver that does not contain cyclic substructure.}
\label{Fig:Tree_Extension_General}
\end{figure}
This argument can be generalized to any tree-like connected Abelian quiver.
More precisely, consider such an Abelian quiver and assume that it has at least $n\geq3$ vertices\footnote{In graph theory, such vertices are of \emph{degree} (or \emph{valency}) equal to $1$.} that are connected only via one $\left(\ell,-k\right)$-edge. For each of these vertices, the charge matrix $\rho$ yields a relation that schematically reads
\begin{align}
    q_{i}\mathcal{X}_{i}=0 \; ,
\end{align}
where $i$ counts the generating elements $\mathcal{X}_{i}$ associated with these outer vertices and $q_{i}$ denotes the out-going charge from the point of view of that vertex. In order to simplify the argument, single out one of these vertices and without loss of generality label it $(1)$ in Figure~\ref{Fig:Tree_Extension_General}, such that all counter-orientated edges are labeled by some charge $-k_j$. For that vertex one has two relations
\begin{subequations}
\begin{align}
    & \ell_{1}\mathcal{X}_1=0 \; , \label{Eq:Tree_Extension_General2}
    \\
    & k_{1}\mathcal{X}_{1}=\sum_{j=2}^{m}p_{j}\mathcal{X}_{j} \; . \label{Eq:Tree_Extension_General1}
\end{align}
\end{subequations}
The latter relation follows from all the edges that are connected to the central vertex in Figure~\ref{Fig:Tree_Extension_General}. Similarly, for each following vertex, one finds relations that schematically read
\begin{align}
    k_{j}\mathcal{X}_{j}=\sum_{t}p_{t}\mathcal{X}_{t} \; .
\end{align}
Multiplying the L.H.S. in \eqref{Eq:Tree_Extension_General1} with $k_{j}$, allows to \emph{move} the R.H.S. to the next layer of the vertices. Since the quiver is a connected tree, one can continue to multiply further factors onto the L.H.S. and by doing so move through each path and each branching of the quiver. Ultimately, one ends in one of the $n$-many degree-$1$ vertices. Multiplying with the remaining charge $q_{i}$ then terminates the R.H.S. such that one obtains
\begin{align}
    \prod_{i\in I}k_{i}\mathcal{X}_{1}=0 \; ,
\end{align}
where the set $I$ labels \emph{all} the vertices of the quiver. Combining this with \eqref{Eq:Tree_Extension_General2} implies 
\begin{align}
    \gcd\left(\ell_{1},\prod_{i\in I}k_{i}\right)\mathcal{X}_{1}=0 \; .
    \label{Eq:Tree_GCD_final}
\end{align}
However, in order for the quiver to be minimal, all embedded chain quivers need to be trivial, i.e.\ the associated $\gcd$-expressions need to trivialize. In particular, this includes the $\gcd$ of $\ell_{1}$ with each counter-orientated charge $k_{i}$ (for $i\in I$). Therefore, the $\gcd$ in \eqref{Eq:Tree_GCD_final} has to be $1$ and the associated hypermultiplet to $\mathcal{X}_{1}$ becomes free and decouples from the theory\footnote{Note that the arguments holds simultaneously for all of the $n$-many degree-$1$ vertices, i.e.\ all $n$-many associated hypermultiplets become free.}, i.e.
\begin{align}
    \mathcal{X}_{1}=0 \; .
\end{align}
To summarize, no connected Abelian tree-like quiver beyond a \emph{pure} chain can be non-trivial and minimal simultaneously. 

\subsection{Remarks on redundancy and higher symmetries}
\label{Sec:AbelianChain_Remarks_Redundancy}
The result of Section \ref{Sec:MinimalAbelianChain} establishes that any chain quiver as shown in Figure \ref{Fig:Complete_Class_AbelianChain_Minimal} is a stable quiver if and only if the conditions \eqref{Eq:AbelianChain_GCD_Condition} are satisfied --- moreover, the minimal Coulomb branch geometry derived to be of type $h_{n,k,\sigma}$, see \eqref{eq:minimalChain_CB_geometry}. Understanding this as a map from (magnetic) quivers to Coulomb branch geometries, this map is certainly many-to-one. In other words, there is redundancy in the gauge theory realization of geometries. Here, this is elaborated on.  

\paragraph{1-form symmetries.}
To begin with, consider a  3-node example (in reduced form) and its mirror
\begin{align}
\mathsf{Q}: \; \raisebox{-.5\height}{
\begin{tikzpicture}
    \tikzset{node distance=2cm};
    \node (1a) at (0,0) [flavour,label={[align=center]below:\small{$1$}\\\small{$(1)$}}] {};
    \node (1b) [gauge,right of=1a,label={[align=center]below:\small{$1$}\\\small{$(2)$}}] {};
    \node (1c) [gauge,right of=1b,label={[align=center]below:\small{$1$}\\\small{$(3)$}}] {};  
    \draw (1a)--(1b);
    \draw[line width=2pt,gray, decoration={markings, mark=at position 0.2 with {\arrow[scale=0.7]{>}}, mark=at position 0.9 with {\arrow[scale=0.7]{<}}}, postaction={decorate}] (1b)to node[midway, above, black] {\small{$(q,\delta)$}} (1c);
\end{tikzpicture}}
\qquad \longleftrightarrow \qquad 
\mathsf{Q}^\vee: \;
\raisebox{-.5\height}{
    \begin{tikzpicture}
    \tikzset{node distance=2cm}; 
    \node (1a) at (0,0) [gauge,label={[align=center]right:$\mathbb{Z}_{\delta}$}] {};
    \node (1b) at (-1,-2) [flavour,label={[align=center]right:\small{$1$}}] {};
    \node (1c) [flavour, right of=1b,label={[align=center]right:\small{$1$}}] {};
    \draw[line width=2pt,gray, decoration={markings, mark=at position 0.5 with {\arrow{>}}}, postaction={decorate}] (1a)to[out=210,in=90] node[below right, black] {\small{$q$}} (1b);
    \draw[line width=1pt,gray] (1a)to[out=330,in=90] (1c);
\end{tikzpicture}
}
\label{eq:3-node_w_1-form}
\end{align}
The quiver $\mathsf{Q}$ has a $\Z_\delta$ 1-form symmetry due to the higher charge $q$ of the hypermultiplet charged under the $\urm(1)_{(3)}$ gauge group factor. However, one can (formally) generate $\mathsf{Q}$ by considering the case $\delta\to1$. Here, $\mathsf{Q}|_{\delta =1}$ degenerates into two copies of $\mathrm{SQED}_1$, while the mirror is simply two free hypermultiplets. Nevertheless, the perspective is useful, as one can imagine $\mathsf{Q}$ to be obtained via gauging a $\Z_\delta$ subgroup of the Coulomb branch isometry group of $\mathsf{Q}|_{\delta=1}$. This is compatible with the behavior of the mirror, wherein a discrete $\Z_\delta$ subgroup of the Higgs branch isometry is gauged --- resulting in $\mathsf{Q}^\vee$. 
The explicit form of the mirror $\mathsf{Q}^\vee$ also manifests the redundancies when focusing on its Higgs branch. Charges $q$ that are equivalent $\bmod \delta$ yield the same Higgs branches.

Now, one can argue that this is the only source of a 1-form symmetry for a stable chain quiver. The explicit parametrization of the reduced stable chain quiver in Figure~\ref{Fig:NVertexChain_Quiver_Reduced_Minimal} already does not admit further 1-form symmetries. To elucidate the underlying reason, consider a 3-node quiver (in reduced form) that exhibits a product 1-form symmetry $\Z_q \times \Z_\delta$:
\begin{align}
\raisebox{-.5\height}{
    \begin{tikzpicture}
    \tikzset{node distance=2cm};
    \node (1a) at (0,0) [flavour,label={[align=center]below:\small{$1$}\\\small{$(1)$}}] {};
    \node (1b) [gauge,right of=1a,label={[align=center]below:\small{$1$}\\\small{$(2)$}}] {};
    \node (1c) [gauge,right of=1b,label={[align=center]below:\small{$1$}\\\small{$(3)$}}] {};
    \draw[line width=2pt,gray, decoration={markings, mark=at position 0.9 with {\arrow[scale=0.7]{<}}}, postaction={decorate}] (1a)to node[midway, above, black] {\small{$(0,q)$}} (1b);
    \draw[line width=2pt,gray, decoration={markings, mark=at position 0.2 with {\arrow[scale=0.7]{>}}, mark=at position 0.9 with {\arrow[scale=0.7]{<}}}, postaction={decorate}] (1b)to node[midway, above, black] {\small{$(q,\delta)$}} (1c);
   \end{tikzpicture}}
   \label{eq:3-node_w_product_1-form}
\end{align}
where one may assume $\gcd(q,\delta)=1$.
In the mirror theory, one consequently expects two discrete gauge group factors\footnote{To see this, observe that \eqref{eq:3-node_w_product_1-form} can be obtained from the $q=1$ case of $\mathsf{Q}$ in \eqref{eq:3-node_w_1-form} by gauging a $\Z_q$ subgroup of the topological symmetry associated with node $(2)$. In the mirror side, there are 2 hypermultiplets of charge $1$ under $\Z_\delta$ gauge group and one gauges a discrete $\Z_q$ subgroup of the flavor symmetry; i.e.\ the first form of $\mathsf{Q}^\vee$ in \eqref{eq:3-node_w_product_1-form_details}.} and the questions is whether there exists a Higgsing with a non-trivial residual theory. It is, however, not hard to see that the unframed quiver form as well as the mirror theory are not stable:
\begin{align}
  \mathsf{Q}:\;  \raisebox{-.5\height}{\begin{tikzpicture}
    \tikzset{node distance=2cm};
    \node (1a) at (0,0) [gauge,label={[align=center]below:\small{$1$}\\\small{$(1)$}}] {};
    \node (1b) [gauge,right of=1a,label={[align=center]below:\small{$1$}\\\small{$(2)$}}] {};
    \node (1c) [gauge,right of=1b,label={[align=center]below:\small{$1$}\\\small{$(3)$}}] {};
    \draw[line width=2pt,gray, decoration={markings, mark=at position 0.2 with {\arrow[scale=0.7]{>}}, mark=at position 0.9 with {\arrow[scale=0.7]{<}}}, postaction={decorate}] (1a)to node[midway, above, black] {\small{$(\delta q,-q)$}} (1b);
    \draw[line width=2pt,gray, decoration={markings, mark=at position 0.2 with {\arrow[scale=0.7]{>}}, mark=at position 0.9 with {\arrow[scale=0.7]{<}}}, postaction={decorate}] (1b)to node[midway, above, black] {\small{$(q,-\delta)$}} (1c);
   \end{tikzpicture}}
   \qquad \longleftrightarrow\qquad
     \mathsf{Q}^\vee:\;
\raisebox{-.5\height}{     \begin{tikzpicture}
    \tikzset{node distance=2cm};
    \node (1a) at (0,0) [gauge,label={[align=center]below:\small{$\Z_\delta$}}] {};
    \node (1b) [gauge,right of=1a,label={[align=center]below:\small{$\Z_q$}}] {};
    \node (1c) [flavour,above of=1a,label={[align=center]right:\small{$1$}}] {};
   \draw (1a)--(1b) (1a)--(1c);
   \end{tikzpicture}}
\cong
\raisebox{-.5\height}{
    \begin{tikzpicture}
    \tikzset{node distance=2cm}; 
    \node (1a) at (0,0) [gauge,label={[align=center]right:$\mathbb{Z}_{q \delta}$}] {};
    \node (1b) at (-1,-2) [flavour,label={[align=center]right:\small{$1$}}] {};
    \node (1c) [flavour, right of=1b,label={[align=center]right:\small{$1$}}] {};
    \draw[line width=2pt,gray, decoration={markings, mark=at position 0.5 with {\arrow{>}}}, postaction={decorate}] (1a)to[out=210,in=90] node[below right, black] {\small{$q$}} (1b);
    \draw[line width=1pt,gray] (1a)to[out=330,in=90] (1c);
\end{tikzpicture}
}
\label{eq:3-node_w_product_1-form_details}
\end{align}
The chain quiver $\mathsf{Q}$ admits a non-trivial decay to a 2-node chain, while the mirror $\mathsf{Q}^\vee$ has a Higgs branch with a non-isolated singularity because the charge of one hypermultiplet and the order of the discrete gauge group are not co-prime.
For a $n$-vertices stable chain quiver, the arguments are analogous. The conclusion is that stable chains have a 1-form symmetry $\Z_{\gcd(l_1,k_{n-1})}$ and that coincides with the gauge group of the mirror theory.

\paragraph{Redundancy.}
Recall that the reduced form of Figure~\ref{Fig:NVertexChain_Quiver_Reduced_Minimal} is achieved by suitably combining the $n$  $\urm(1)$ gauge group factors of the $n$-vertex chain. However, there is still redundancy left in this description: it originates from permuting the $n-1$ hypermultiplets as well as sign flips of the $\urm(1)$ charges. Thus, in the reduce form of an $n$-vertex chain, there is an $S_{n-1}$ permutation group interchanging the hypermultiplets and a $\Z_2^{n-2}$ product group that flips the charges.

This can be illustrated with a Hilbert series computation for the quiver \eqref{eq:3-node_w_1-form}; for which one finds
\begin{align}
    \mathrm{HS}=
    \frac{\sum_{n=0}^{\delta-1} \left(t^n -t^{a-n}\right) \left(t^{qn \bmod \delta }  + t^{\delta -(q n \bmod \delta)}
    \right)}{(1-t^2)^2 (1-t^\delta)^2} \,.
\end{align}
Here, $n=3$ and one expects an $\Z_2\times S_2 \cong \Z_2^2$ redundancy. Inspecting for examples $\delta=7$, one finds the same moduli space geometry for $q=2,3,4,5$. This is interpreted as orbit of size $4$ under the redundancy group.

\FloatBarrier

\section{Cycle quivers with minimal Coulomb branch}
\label{Sec:Abelian_Cycle}
\begin{figure}[t!]
\begin{subfigure}[t]{1\textwidth}
    \centering
    \includegraphics[page=1]{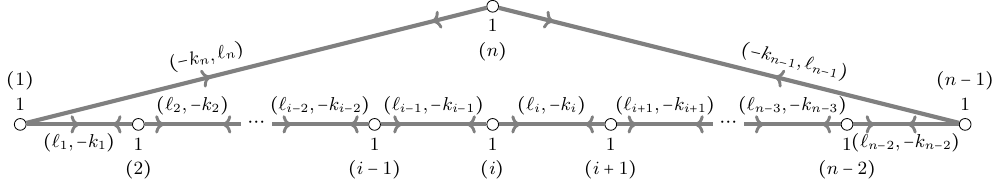}
\caption{}
\label{Fig:Generic_Cycle}
\end{subfigure}
\begin{subfigure}[t]{1\textwidth}
    \centering
    \includegraphics[page=2]{Abelian_Decay_Fission_Fig15.pdf}
\caption{}
\label{Fig:Generic_Cycle_Minimal_Mirror}
\end{subfigure}
\caption{\subref{Fig:Generic_Cycle}: generic $n$-vertex Abelian cycle quiver. The vertex labeled $\left(1\right)$ is in the lower left corner; the remaining vertices and edges are labeled with respect to \emph{moving} from this vertex through the entire cycle, i.e.\ the charges $\ell_{i}$ are always orientated away from the vertex $\left(i\right)$, and charges denoted $k_{i}$ are always orientated towards the vertex $\left(i\right)$. The cycle exhibits a minimal Coulomb branch if it satisfies the constraints \eqref{Eq:Cycle_Minimal_Conditions}, in which case its mirror dual is given by \subref{Fig:Generic_Cycle_Minimal_Mirror}: a SQED-like theory with $n$-many hypermultiplets with charges as in \eqref{eq:defSigma}.}
\label{Fig:NCycleQuiver_Minimal}
\end{figure}

In this section the class of 3d $\Ncal=4$ Abelian $n$-vertices quiver gauge theories with a cyclic (sub\mbox{-})structure and well-defined length that have a minimal Coulomb branch are defined. 
A cycle of the form shown in Figure \ref{Fig:Generic_Cycle} is said to have a \emph{well-defined length} if and only if 
\begin{equation}
    \prod_{i \in \mathbb{Z}_n} \ell_i = \prod_{i \in \mathbb{Z}_n}  k_i \,. 
\label{eq:length_def}
\end{equation}
Any other cycle is said to not admit such a well-defined length.\footnote{This concept is defined in analogy with non-simply laced Dynkin diagrams, where the nodes denote roots, and the lacedness encodes ratios of squared lengths for the roots. } 

Building upon  Section~\ref{Sec:Abelian_Chain}, the class of minimal $n$-vertices Abelian cycle theories --- $n$-vertices \emph{cycle quivers} for short --- of the form shown in Figure~\ref{Fig:Generic_Cycle} with a well-defined length is discussed and characterized in Section~\ref{Sec:Abelian_Cycle_Minimal}. Following that, Section~\ref{Sec:Abelian_Cycle_Extension} contains some remarks on cyclic theories without a well-defined length and extended Abelian quivers with cyclic sub-structure.

\subsection{Stable Abelian cycle quivers}
\label{Sec:Abelian_Cycle_Minimal}

Here, the focus is solely on cycle quivers with well-defined length. 

\paragraph{Minimal condition using Decay and Fission.}
Using the decay and fission algorithm one can construct the schematic Hasse diagram for the Coulomb branch stratification of the generic Abelian $n$-vertices cycle quiver of Figure~\ref{Fig:Generic_Cycle}; each decay channel yields embedded $m$-vertices chains $\left(2\leq m\leq n-1\right)$ and products of these. Therefore, the minimal $n$-vertices Abelian cycle can be identified as the quiver that only contains trivially embedded $m$-vertices chains. This is encoded in the following constraints:
\begin{align}
\label{Eq:Cycle_Minimal_Conditions}
    \boxed{    \gcd\left(\ell_{i},k_{j}\right)=1 \quad \text{for all} \, \left(i,j\right) \in \mathbb{Z}_n^2  \quad \text{such that} \quad i-j \not\equiv 1,2 \, \text{mod} \, n } \; .
\end{align}

\paragraph{Proof using 3d mirror symmetry.}
For a generic $n$-vertices cycle quiver as in Figure \ref{Fig:Generic_Cycle} the charge matrix $\rho: \Z^{n}\mapsto\Z^{n}$ reads 
\begin{equation}
\rho=
\begin{tikzpicture}[baseline]
\matrix (m) [matrix of math nodes,nodes in empty cells,right delimiter={)},left delimiter={(} ]{
\ell_1 & -k_1 & 0 & 0 & & & & & 0 \\
0 & \ell_2 & -k_2 & 0 & & & & & 0 \\
0 & 0 & \ell_3 & -k_3 & & & & & 0 \\
& & & & & & & &\\
& & & & & & & &\\
& & & & & & & &\\
0 & & & & & & \ell_{n-2} & -k_{n-2} & 0 \\
0 & & & & & & 0 & \ell_{n-1} & -k_{n-1} \\
-k_{n} & & & & & & 0 & 0 & \ell_{n} \\
} ;
\draw[loosely dotted, thick] (m-1-4)-- (m-1-9);
\draw[loosely dotted, thick] (m-3-9)-- (m-7-9);
\draw[loosely dotted, thick] (m-3-1)-- (m-7-1);
\draw[loosely dotted, thick] (m-9-1)-- (m-9-7);
\draw[loosely dotted, thick] (m-3-3)-- (m-7-7);
\end{tikzpicture} \; . \label{Eq:NCycle_Charge}
\end{equation}
Note that \eqref{eq:length_def} is equivalent to $\det \rho = 0$. In Appendix~\ref{app:mirror_cycle_reduce}, we prove the following lemma: 

\emph{\underline{Lemma}: Under the assumptions \eqref{eq:length_def} and \eqref{Eq:Cycle_Minimal_Conditions}, the 3d mirror theory of the quiver in Figure \ref{Fig:Generic_Cycle} is the $\urm(1)$ gauge theory with $n$ hypermultiplets of charges 
\begin{equation}
   \boxed{ \sigma_i := \gcd (\ell_{i+2} , k_i) \qquad \text{for} \qquad i \in \mathbb{Z}_n  } \, , 
    \label{eq:defSigma}
\end{equation}
and the Coulomb branch of Figure \ref{Fig:Generic_Cycle} is therefore $\overline{h}_{n-1, \sigma}$. 
}

Using similar arguments, one can also show that if one of the conditions in \eqref{Eq:Cycle_Minimal_Conditions} is violated, then the 3d mirror theory has gauge group which is a product, and the Coulomb branch is not minimal.

\begin{figure}[t!]
    \centering
    \includegraphics[page=1]{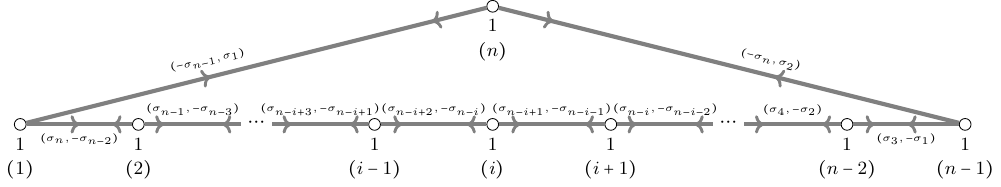}
\caption{Mirror cycle quiver for the SQED theories considered by Namikawa. It is apparent that all conditions \eqref{Eq:Cycle_Minimal_Conditions} are satisfied.}
\label{Fig:Namikawa}
\end{figure}

\paragraph{Relation to Namikawa theories.}
Note that \eqref{Eq:Cycle_Minimal_Conditions} implies that the $\sigma_i$ defined in \eqref{eq:defSigma} are pairwise coprime, i.e.\  $\gcd(\sigma_i,\sigma_j)=1$ for all $i \neq j$. 
In \cite{namikawa2023remark}, the hyper-K\"ahler quotient construction for $\urm(1)$ with $n$ hypermultiplets of pairwise co-prime charges $(\sigma_1,\sigma_2,\ldots,\sigma_n)$ has been considered and shown to yield an isolated symplectic singularity. Applying the cokernel construction for the 3d $\Ncal=4$ mirror theories yields the following charge matrix
\begin{align}
\label{eq:charge_matrix_Namikawa_mirror}
 \begin{pmatrix}
        \sigma_{n-1}  & -\sigma_1    & 0    & 0       & \cdots & 0        & 0        &  0    & 0 \\
        0    & \sigma_n     & -\sigma_2 & 0    & \cdots & 0        & 0 &  0 &  0\\
        0    & 0       & \sigma_1  & -\sigma_3  & \cdots & 0 & 0  &  0       &  0\\
        0    & 0       & 0    & \sigma_2     & \cdots & 0  & 0        &  0       &  0\\
     \vdots  & \vdots  & \vdots & \vdots &        & \vdots   & \vdots   & \vdots & \vdots\\
        0    &   0     & 0    & 0     & \cdots & \sigma_{n-5}  & -\sigma_{n-3} & 0        & 0\\
        0    &   0     & 0    & 0     & \cdots & 0        & \sigma_{n-4}  & -\sigma_{n-2} &  0\\
        0    &   0     & 0    & 0      & \cdots & 0       & 0        & \sigma_{n-3}  & -\sigma_{n-1} \\
        -\sigma_{n}    &   0     & 0    & 0     & \cdots & 0     & 0        & 0        & \sigma_{n-2} \\
    \end{pmatrix}
\end{align}
which leads to the cycle quiver theory shown in Figure \ref{Fig:Namikawa} with a well-defined length. This is compatible with \eqref{eq:defSigma} and \eqref{Eq:NCycle_Charge}.

\subsection{Extended Abelian cycle quivers and ill-defined length}
\label{Sec:Abelian_Cycle_Extension}

In this Section the connection between the kernel of the gauge group representation of a cycle quiver and the property of having a well-defined length is commented on.

To begin with, consider the representation $\gamma$ of the gauge group acting on the hypermultiplets of the Abelian $n$-vertices cycle quiver as in Figure~\ref{Fig:Generic_Cycle}
\begin{align}
    \gamma:&\;  
         \urm(1)^n \to  
   \; \mathrm{End}(\mathbb{H}^n) \\
   &\left(g_{1},g_{2},\dots,g_{i},\dots,g_{n}\right)
   \mapsto\; \mathrm{diag}\left(g_{1}^{\ell_{1}}g_{2}^{-k_{1}},g_{2}^{\ell_{2}}g_{3}^{-k_{2}},\dots,g_{i}^{\ell_{i}}g_{i+1}^{-k_{i}},\dots,g_{n-1}^{\ell_{n-1}}g_{n}^{-k_{n-1}},g_{n}^{\ell_{n}}g_{1}^{-k_{n}}\right) \notag
\end{align}
and its kernel\footnote{Choosing representatives of these equivalence classes when ungauging the (connected part of the) kernel defines a specific choice of framing, which does not need to be the complete ungauging on one vertex only.}
\begin{align}
    \mathrm{ker}\left(\gamma\right)=\left\{\left(g_{1},g_{1}^{\tfrac{\ell_{1}}{k_{i}}},g_{1}^{\frac{\ell_{1}\ell_{2}}{k_{1}k_{2}}},\ldots,g_{1}^{\tfrac{\prod_{m=1}^{i}\ell_{m}}{\prod_{n=1}^{i}k_{n}}},\ldots,g_{1}^{\tfrac{k_{n}}{\ell_{1}}}\right)\in\urm(1)^{n} \; \bigg| \; \text{s.t. } \; g_{1}^{\prod_{i=1}^{n}\ell_{i}}\stackrel{!}{=}g_{1}^{\prod_{j=1}^{n}k_{j}}\right\} \; .
\end{align}
There are the following three distinct cases: 
\begin{compactenum}[(i)]
    \item \label{comp:1} There exist a well-defined length, i.e.\ $\prod_{i=1}^{n}\ell_{i}=\prod_{j=1}^{n}k_{j}$. In that case, the kernel contains a continuous factor which can be identified as a diagonally acting $\urm(1)$ (note that there can also be discrete factors if \eqref{Eq:Cycle_Minimal_Conditions} is not fulfilled). It is usually assumed that one has such a kernel and the usual framing prescription refers to un-gauging it completely on one vertex. However, we see in this example that this requires well-defined length. 
    \item \label{comp:2} Another possibility is $\prod_{i=1}^{n}\ell_{i}\neq\prod_{j=1}^{n}k_{j}$, for which one only finds a discrete kernel of the gauge action. In particular, one finds $g_{1}\in\Z_{|\prod_{i=1}^{n}\ell_{i}-\prod_{j=1}^{n}k_{j}|}$.
    \item \label{comp:3} In particular, for $\prod_{i=1}^{n}\ell_{i}=\prod_{j=1}^{n}k_j \pm1$, the kernel is trivial.
\end{compactenum}
Based on this, one can now motivate the restriction to cycle quivers with well-defined length: If a cycle quiver does not have a well-defined length, it is either a trivial theory\footnote{The relation defined in \ref{comp:3} implies that all the hypermultiplet in the mirror dual of the initial cycle quiver decouple and become free. This can be deduced from the relations that define the cokernel of the charge matrix \eqref{Eq:NCycle_Charge}.} or its mirror theory is equipped with a discrete gauge group only. In case of the latter, it stands to reason that if the cycle quiver has a minimal Coulomb branch, it can equivalently be described by a minimal chain quiver. 

\begin{figure}[t!]
\begin{subfigure}[t]{0.5\textwidth}
    \centering
    \includegraphics[page=1]{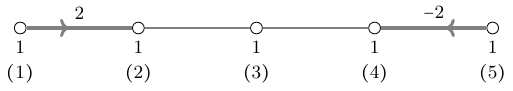}
\caption{}
\label{Fig:Extended_Cycle_Example_1}
\end{subfigure}
\begin{subfigure}[t]{0.35\textwidth}
    \centering
    \includegraphics[page=2]{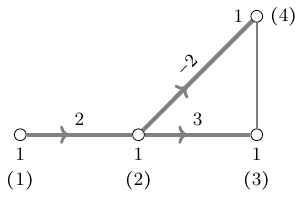}
\caption{}
\label{Fig:Extended_Cycle_Example_2}
\end{subfigure}
\begin{subfigure}[t]{0.125\textwidth}
    \centering
    \includegraphics[page=3]{Abelian_Decay_Fission_Fig17.pdf}
\caption{}
\label{Fig:Extended_Cycle_Example_3}
\end{subfigure}
\caption{An Abelian chain quiver (Figure \subref{Fig:Extended_Cycle_Example_1}) and an extended cycle quiver (Figure \subref{Fig:Extended_Cycle_Example_2}) that have the same 3d $\Ncal=4$ mirror dual (Figure \subref{Fig:Extended_Cycle_Example_3}).
}
\label{Fig:Extended_Cycle_Example}
\end{figure}

Furthermore, an Abelian quiver with cyclic sub-structures beyond the \emph{pure} cycle quiver can only be minimal if none of the cyclic sub-structures have a well-defined length, since such cycle quiver are always good theories in the physical sense\footnote{Recall from before that for the mirror dual to a cycle theory with well-defined length all hypermultiplets are coupled to a $\urm(1)$ gauge group, thus no free hypermultiplets. This holds true irregardless of having a minimal Coulomb branch (equivalently, minimal Higgs branch for the mirror theory) or not.}, and, therefore, realized decay or fission products in the Hasse diagram of the overall quiver theory. However, if the cyclic sub-structures do not have a well-defined length, the overall quiver theory has a discrete or trivial kernel for the gauge group action; thus, having either a discrete gauge group in its mirror dual or being a trivial theory. In case of the former, the motivated assumption is again that if the theory has a minimal Coulomb branch, that it can be equivalently described by a stable chain quiver. To illustrate the argument, consider the specific example\footnote{Note that this example is also used as Example 1 in Section \ref{Sec:Quiver_reduced_form}.} showcased in Figure \ref{Fig:Extended_Cycle_Example}.
The charge matrices of the chain quiver in Figure~\ref{Fig:Extended_Cycle_Example_1} and the extended cycle quiver in Figure~\ref{Fig:Extended_Cycle_Example_2} read
\begin{align}
        \begin{pmatrix}
        2 & -1 & 0 & 0 & 0\\
        0 & 1 & -1 & 0 & 0\\
        0 & 0 & 1 & -1 & 0\\
        0 & 0 & 0 & 1 & -2\\
        \end{pmatrix} \; , \quad \begin{pmatrix}
        2 & -1 & 0 & 0 \\
        0 & 3 & -1 & 0 \\
        0 & 0 & 1 & -1 \\
        0 & -2 & 0 & 1 \\
        \end{pmatrix} \; ,
\end{align}
respectively. The 3d $\Ncal=4$ mirror dual to the chain quiver is defined by the generating set of elements $\left\{\mathcal{X}_{i}\right\}_{i=1}^{4}$, satisfying the relations
\begin{align}
    2\mathcal{X}_{1}=0, \quad \mathcal{X}_{1}=\mathcal{X}_{2}, \quad \mathcal{X}_{2}=\mathcal{X}_{3}, \quad \mathcal{X}_{3}=\mathcal{X}_{4}, \quad 2\mathcal{X}_{4}=0 \; ,
\end{align}
and the mirror dual to the cycle quiver by the generating set of elements $\left\{\mathcal{Y}\right\}_{i=1}^{4}$, with the relations
\begin{align}
    2\mathcal{Y}_{1}=0, \quad \mathcal{Y}_{1}=3\mathcal{Y}_{2}-2\mathcal{Y}_{4}, \quad \mathcal{Y}_{2}=\mathcal{Y}_{3}, \quad \mathcal{Y}_{3}=\mathcal{Y}_{4} \; .
\end{align}
The two mirror theories can be identified as the same orbifold theory of Figure \ref{Fig:Extended_Cycle_Example_3}, thus, the chain quiver and extended cycle quiver can be identified as describing equivalent physical theories.

\subsection{Remarks on redundancy and higher symmetries}

The minimality conditions \eqref{Eq:Cycle_Minimal_Conditions} leave two set of integers unconstrained, namely 
\begin{equation}
    \sigma_i := \gcd (\ell_{i+2} , k_i) \qquad \textrm{and} \qquad \gcd (\ell_{i+1} , k_i) \, . 
\end{equation}
The charge assignment for the geometry \eqref{eq:defSigma} only uses the $\sigma_i$; it is natural to wonder about the impact of $\gcd(\ell_{i+1},k_i)$. Here, their role is illuminated. Again, these GCDs showcase the issue of redundancy in the map quiver to geometries, and, in the case of non-minimal cycles, are a source of 1-form symmetries.

\paragraph{1-form symmetries.}
Consider the following $6$-vertex cycle quiver with well-defined length and two non-trivial $\gcd(\ell_{i+1},k_i)$:
\begin{align}
\mathsf{Q}: \quad
\raisebox{-.5\height}{
    \begin{tikzpicture}
    \tikzset{node distance=1.5cm};
    \node (20) at (10.5,1.5) [gauge,scale=0.7,label=above:\small{$1$},label=above left:\small{$(6)$}] {};
    \node (2a) at (7.5,0) [gauge,scale=0.7,label={[align=center]below:\small{$1$}\\\small{$(1)$}}] {};
    \node (2b) [gauge,right of=2a,scale=0.7,label={[align=center]below:\small{$1$}\\\small{$(2)$}}] {};
    \node (2c) [gauge,right of=2b,scale=0.7,label={[align=center]below:\small{$1$}\\\small{$(3)$}}] {};
    \node (2d) [gauge,right of=2c,scale=0.7,label={[align=center]below:\small{$1$}\\\small{$(4)$}}] {};
    \node (2e) [gauge,right of=2d,scale=0.7,label={[align=center]below:\small{$1$}\\\small{$(5)$}}] {};
    \draw[line width=2pt,gray, decoration={markings, mark=at position 0.5 with {\arrow{>}}}, postaction={decorate}] (20)to node[midway, above left, black] {\small{$q$}} (2a);
    \draw[line width=2pt,gray, decoration={markings, mark=at position 0.5 with {\arrow{>}}}, postaction={decorate}] (20)to node[midway, above right, black] {\small{$-q$}} (2e);
    \draw[line width=2pt,gray] (2a) to (2b);
    \draw[line width=2pt,gray, decoration={markings, mark=at position 0.5 with {\arrow{<}}}, postaction={decorate}] (2b)to node[midway, below, black] {\small{$-p$}} (2c);
    \draw[line width=2pt,gray, decoration={markings, mark=at position 0.5 with {\arrow{>}}}, postaction={decorate}] (2c)to node[midway, below, black] {\small{$p$}} (2d);
    \draw[line width=2pt,gray] (2d) to (2e);
\end{tikzpicture}}
\label{eq:chain_example_1-from}
\end{align}
and one observes that depending on $\gcd(p,q)$ the chain can be stable or not. To illustrate, recall the cokernel is described by the relations :
\begin{align}
    \Xcal_6=\Xcal_1 =\Xcal_2
    \;,\quad
      \Xcal_3 =\Xcal_4 =\Xcal_5
      \;,\quad
    p(\Xcal_5 -\Xcal_6)= 0  
    \;,\quad
    q(\Xcal_5 -\Xcal_6)=0
    \;. \label{eq:cokernel_example}
\end{align}
For $\gcd(p,q)=1$, there is only the $\urm(1)$ gauge group in the mirror, while for $\gcd(p,q)=\delta>1$ the mirror gauge group is $\urm(1)\times \Z_\delta$.  Computing the charges one arrives at:
\begin{align}
  \mathsf{Q}^\vee_{\gcd(p,q)=1}:\quad   \raisebox{-.5\height}{
    \begin{tikzpicture}
    \tikzset{node distance=1.5cm}; 
    \node (1a) at (0,0) [gauge,label={[align=center]below:$1$}] {};
    \node (1b) [flavour,right of=1a,label={[align=center]below:\small{$6$}}] {};
    \draw[line width=1pt,gray] (1a)to (1b);
\end{tikzpicture}
}\qquad \qquad
  \mathsf{Q}^\vee_{\gcd(p,q)=\delta>1}:\quad   \raisebox{-.5\height}{
    \begin{tikzpicture}
    \tikzset{node distance=2cm}; 
    \node (1a) at (0,0) [gauge,label={[align=center]below:$1$}] {};
    \node (1b) [gauge,left of=1a,label={[align=center]below:$\mathbb{Z}_{\delta}$}] {};
    \node (1c) [flavour, right of=1a,label={[align=center]below:\small{$3$}}] {};
    \draw[line width=1pt,gray] (1a)to (1b);
     \draw[line width=1pt,gray,transform canvas={xshift=0pt,yshift=2.5pt}](1a)--(1b);
      \draw[line width=1pt,gray,transform canvas={xshift=0pt,yshift=-2.5pt}](1a)--(1b);
    \draw[line width=1pt,gray] (1a)to (1c);
\end{tikzpicture}
} \label{eq:mirror_chain_example_1-from}
\end{align}
and the corresponding Higgs branch is only isolated for $\delta=1$.
Of course, this is fully transparent by the decay and fission argument, as the non-trivial $\gcd$ enables a decay to embedded stable chain quivers. Note in particular, that the $p$ and $q$ values disappear in the mirror as long as $\delta=1$, which showcases the redundancy inherent to the $\gcd(\ell_{i+1},k_i)$.

In a vein similar to Section \ref{Sec:AbelianChain_Remarks_Redundancy}, the mirror \eqref{eq:mirror_chain_example_1-from} for $\delta>1$ exhibits a 1-form symmetry, whose origin can be tracked in \eqref{eq:chain_example_1-from}. In fact, it is more convenient to work with a framed quiver version; which for $\gcd(p,q)=\delta$ reads
\begin{align}
\mathsf{Q}: \quad
\raisebox{-.5\height}{
    \begin{tikzpicture}
    \tikzset{node distance=1.5cm};
    \node (2a) at (7.5,0) [gauge,scale=0.7,label={[align=center]below:\small{$1$}}] {};
    \node (2b) [gauge,right of=2a,scale=0.7,label={[align=center]below:\small{$1$}}] {};
    \node (2c) [gauge,right of=2b,scale=0.7,label={[align=center]below:\small{$1$}}] {};
    \node (2d) [gauge,right of=2c,scale=0.7,label={[align=center]below:\small{$1$}}] {};
    \node (2e) [gauge,right of=2d,scale=0.7,label={[align=center]below:\small{$1$}}] {};
    \node (f1) [flavour, above of=2a,label={left:$1$}] {};
    \node (f2) [flavour, above of=2e,label={right:$1$}] {};
    \draw[line width=1pt] (f1) to (2a);
    \draw[line width=1pt] (f2) to (2e);
    \draw[line width=2pt,gray] (2a) to (2b);
    \draw[line width=2pt,gray, decoration={markings, mark=at position 0.5 with {\arrow{<}}}, postaction={decorate}] (2b)to node[midway, below, black] {\small{$-\delta$}} (2c);
    \draw[line width=2pt,gray, decoration={markings, mark=at position 0.5 with {\arrow{>}}}, postaction={decorate}] (2c)to node[midway, below, black] {\small{$\delta$}} (2d);
    \draw[line width=2pt,gray] (2d) to (2e);
\end{tikzpicture}} \,.
\end{align}
Again, for $\delta>1$ there is an apparent 1-form symmetry due to the trivially acting $\Z_q$ center gauge symmetry. The connection between the $\delta=1$ and $\delta>1$ scenario is realized by discrete gauging of global symmetries, cf.\ \cite{Nawata:2023rdx}. In contrast to the stable chain quivers\footnote{For the stable $n$-vertex chain, the degenerate starting point is the flat quaternions $\mathbb{H}^{n-1}$. Hence, discretely gauging a $\Z_\delta$ can produce non-trivial and isolated singularities; see Section \ref{Sec:AbelianChain_Remarks_Redundancy}. This is the reason 1-form symmetries exist in stable chains, but not in stable cycles. }, such gaugings in the cycle case cannot yield a stable quiver because the starting point (i.e.\ the $q=1$ case in the example) is already a non-trivial and minimal geometry. Nonetheless, the $\gcd(\ell_{i+1},k_i)$ are possible sources of 1-form symmetries in non-stable cycles. 

Therefore, this example can then be turned into an argument about 1-from symmetries valid for any stable cycle quiver with well-defined length.  Consider the cycle of Figure \ref{Fig:Generic_Cycle}. To obtain a cyclic group $\Z_\delta$ in the cokernel (and hence a discrete $\Z_\delta$  gauge group factor on the mirror) one needs to have two nodes $(i)$ and $(j)$ with charges satisfying 
\begin{align}
    \gcd(\ell_i,k_{i-1}) = g_{i-1}>1
    \;,\qquad
    \gcd(\ell_j,k_{j-1}) = g_{j-1}>1
    \;,\qquad
    \gcd(g_{i-1},g_{j-1}) \equiv \delta>1 \,.
\end{align}
This is because one then gets a set of conditions analogous to \eqref{eq:cokernel_example}.
But this necessarily violates the minimality conditions, as $\gcd(\ell_i,k_{j-1}) \geq \delta$ and $ \gcd(\ell_j,k_{i-1})\geq \delta$. Hence, there exists at least one embedded non-trivial chain quiver and the cycle quiver is not stable. Consequently, a stable cycle quiver cannot exhibit any 1-from symmetry and, equivalently, the mirror can be described by a SQED-type theory.

\paragraph{Redundancy.}
The above argument then shows again a large redundancy when mapping quivers to geometries. Given a cycle quiver that satisfies the minimality conditions \eqref{Eq:Cycle_Minimal_Conditions}, one can remove the $\gcd(\ell_{i+1},k_i)$ of the charges at each gauge node $(i+1)$. This then should define some ``reduced'' version of the cycle. One still expects the usual redundancies, stemming from permuting the hypermultiplets and flipping the signs of the hypermultiplet charges, to show up.

\section{Conclusion and Outlook} \label{Sec:conclusion}
Abelian 3d $\Ncal=4$ quiver gauge theories and their mirror symmetry properties have long been subject of research, both in physics and mathematics. However, it appears that only the recent appreciation of global forms of symmetries and gauge groups has paved the way for studying discrete group factors.

A first result of this paper, perhaps a byproduct, is using the cokernel method of Section~\ref{Sec:Mirror_Construction} to compute mirror dual theories with discrete gauge group factors. 

The second result concerns tree-like Abelian quiver discussed in Section \ref{Sec:Abelian_Chain}. Here, the chain quivers play a pivotal role. Under certain conditions \eqref{Eq:AbelianChain_GCD_Condition}, their Coulomb branches are minimal, in fact of type $h_{n,k,\sigma}$, and the theories are stable quivers. Moreover, among the tree-like quivers, only chain quivers can be stable quiver. No other extended quiver tree structure yields a minimal Coulomb branch.   

The third result is within the realms of Abelian quiver theories that contain a cycle, as discussed in Section \ref{Sec:Abelian_Cycle}. For pure cycle quivers with well-defined length, the Coulomb branches are minimal, of type $ \overline{h}_{n,\sigma}$, provided the conditions \eqref{Eq:Cycle_Minimal_Conditions} are satisfied; the main argument hinges on the decay and fission algorithm. Any extension that contains such a cycle with well-defined length is shown to automatically be non-minimal. Moreover, pure cycle quivers that do not have a well-defined length are argued to have Coulomb branches of type $h_{n,k,\sigma}$; thus, being equivalent to stable chain quivers.

These outcomes establish a classification of isolated conical symplectic singularities that can be realized as Coulomb branches of 3d $\Ncal=4$ Abelian quiver gauge theories. Of course, the stable chain quiver of Figure \ref{Fig:Complete_Class_AbelianChain_Minimal} and the stable cycle quiver of Figure \ref{Fig:Generic_Cycle} vastly over-parametrize the encountered ICSS geometries. This is, however, not unexpected as gauge theories intrinsically contain redundancies. Moreover, the given parametrization is, in fact, a crucial input for the decay and fission algorithm and its generalizations. To see this, recall that in order to trace out the Hasse diagram for a given (magnetic) quiver the adjacency matrix (and hence the charge matrix) remains untouched, while the only decaying of fissioning quantity is the rank vector.

\paragraph{Acknowledgments.}
The authors thank Julius Grimminger and Fabio Marino for stimulating discussions. The work of SMS and MS is supported by the Austrian Science Fund (FWF), START project ``Phases of quantum field theories: symmetries and vacua'' STA 73-N [grant DOI: 10.55776/STA73]. SMS and MS also acknowledge support from the Faculty of Physics, University of Vienna. AB and MS thank the MITP workshop on ``Geometry and Symmetries of SCFTs'' for hospitality. The work of QL is supported by \'Ecole Normale Supérieure - PSL through a CDSN doctoral grant.

\appendix

\section{Matching the mirror dual for the stable chain}
\label{Sec:Appendix_Mirror_Matching}

Recall for the $n$-vertices Abelian chain quiver with minimal Coulomb branch the relations defining the cokernel of the charge matrix (see \eqref{Eq:General_AbelianChain_Relations_Charge} to \eqref{Eq:General_AbelianChain_Relations_GCD}):
\begin{alignat}{2}
    k_{i}\mathcal{X}_{i}&=\ell_{i+1}\mathcal{X}_{i+1} &\qquad& \forall \, i: \, 1\leq i\leq n-2 \;, \label{Eq:Appendix_Mirror_Matching_Minimal_Relation}
    \\
    \gcd\left(\ell_{1},k_{n-1}\right)\mathcal{X}_{i}&=0 &\qquad& \forall \, i:\, 1\leq i\leq n-1 \;.
\end{alignat}
One can invert the relations \eqref{Eq:Appendix_Mirror_Matching_Minimal_Relation} to perform a basis change such that all hypermultiplet charges are fixed by assigning one charge only. Recall from Section \ref{Sec:Abelian_Chain_Equivalence} the decomposition (see \eqref{Eq:Bezout_coeffs_1} and \eqref{Eq:Bezout_coeffs_2}) of the $\gcd$-function via a choice of Bézout coefficients $u_{i}$ and $\upsilon_{i}$, in particular:
\begin{subequations}
\begin{alignat}{2}
\gcd\left(\prod_{j=1}^{i}\ell_{j},k_{i},\right)&=u_{i}\prod_{j=1}^{i}\ell_{j}-\upsilon_{k}k_{i}=1 &\qquad& \forall \, i: \, 1\leq i\leq n-2 \; , \label{Eq:Appendix_Mirror_Matching_Bezout}
    \\
    q_{j}&=\ell_{j}\upsilon_{j-1} &\qquad& \forall \, i: \, 2\leq j\leq n-1
\end{alignat}
\end{subequations}
Solving the relations
\begin{align}
    -\upsilon_{k}k_{i}\mathcal{X}_{i}=\ell_{i+1}\mathcal{X}_{i+1} \qquad \forall \, i: \, 1\leq i\leq n-2
\end{align}
using \eqref{Eq:Appendix_Mirror_Matching_Bezout} and \eqref{Eq:General_AbelianChain_Relation_Mod0}, yields
\begin{align}
    \mathcal{X}_{i}=(-1)^{n-1-i}\prod_{j=i+1}^{n-1}q_{j}\mathcal{X}_{n-1} \qquad \forall \, i: \, 1\leq i\leq n-2 \; .
\end{align}
The charge assignments for the $\Ncal=2$ chiral pairs $\{\mathcal({A}_{i},\widetilde{\mathcal{A}_{i}})\}_{i}^{n-1}$ associated with the generating elements $\left\{\mathcal{X}_{i}\right\}_{i=1}^{n-1}$ reads
\begin{align}
    \left[\mathcal{A}_{n-1}\right]=1 \quad \land \quad \left[\mathcal{A}_{i}\right]=(-1)^{n-1-i}\prod_{j=i+1}^{n-1}q_{j} \quad \forall\, i:\, 1\leq i\leq n-2 \;.
\end{align}
Note that the charges depend on the choice of the Bézout coefficients $\upsilon_{i}$, and that this is the mirror dual theory (see Figure \ref{Fig:NVertexChain_Quiver_Reduced_Mirror}) to the $n$-vertices Abelian chain with minimal Coulomb branch in the reduced representation.

Indeed, there is no non-trivial higgsing possible in the mirror dual:
\begin{align}
    \gcd\left(\prod_{j=i+1}^{n-1}\upsilon_{j-1}\ell_{j},\gcd\left(\ell_{1},k_{n-1}\right)\right) & \stackrel{\phantom{\eqref{Eq:AbelianChain_GCD_Condition2}}}{=}\gcd\left(\gcd\left(\prod_{j=i+1}^{n-1}\upsilon_{j-1}\ell_{j},k_{n-1}\right),\ell_{1}\right)
    \\
    & \stackrel{\eqref{Eq:AbelianChain_GCD_Condition2}}{=}\gcd\left(\gcd\left(\prod_{j=i+1}^{n-1}\upsilon_{j-1},k_{n-1}\right),\ell_{1}\right) \nonumber
    \\
    & \stackrel{\phantom{\eqref{Eq:AbelianChain_GCD_Condition2}}}{=}\gcd\Bigg(k_{n-1},\underbrace{\gcd\left(\prod_{j=i+1}^{n-1}\upsilon_{j-1},\ell_{1}\right)}_{=1}\Bigg)=1 \nonumber \; .
\end{align}
For the last equality consider for all $j$ with $ 1\leq j\leq n-2$
\begin{align}
    \gcd\left(\ell_{1},\upsilon_{j}\right) &\stackrel{\eqref{Eq:AbelianChain_GCD_Condition1}}{=}\gcd\left(\ell_{1},k_{j}\upsilon_{j}\right)
    \\
    &\stackrel{\eqref{Eq:Bezout_coeffs_1}}{=}\gcd\left(\ell_{1},u_{j}\left(\prod_{m=1}^{j}\ell_{m}\right)-1\right) \nonumber
    \\
    &\stackrel{\phantom{\eqref{Eq:Bezout_coeffs_1}}}{\, \equiv}\gcd\left(\ell_{1},U\ell_{1}-1\right)=1 \qquad \forall \, \ell_{1}\in\Z_{\lessgtr0} \; . \nonumber
\end{align}

\section{Mirror of the cycle quiver}
\label{app:mirror_cycle_reduce}

We want to compute the 3d mirror of the quiver defined by the charge matrix \eqref{Eq:NCycle_Charge} satisfying the assumptions \eqref{eq:length_def} and \eqref{Eq:Cycle_Minimal_Conditions}. The strategy is to find a Smith decomposition of the charge matrix: the Smith normal form gives the mirror gauge group, and the charges of hypermultiplets can be read from the invertible integer matrices involved in the Smith decomposition. The first step is to follow the same route as in the Example 2 in Section~\ref{Sec:Quiver_reduced_form}. 

To this end, we need some notations. 
The fact that for all $i=1,\dots,n-2$, \eqref{Eq:Cycle_Minimal_Conditions} implies that $k_i$ is coprime with $\ell_1 \cdots \ell_i$ means that we can find integers $u_i$ and $v_i$ such that 
\begin{equation}
   \left( \prod\limits_{j=1}^i \ell_j \right) u_i - k_i v_i = 1 \, . 
\end{equation}
Also, call $\delta := \gcd (\ell_1 , k_{n-1})$. Using \eqref{Eq:Cycle_Minimal_Conditions} again, we can find integers $u_{n-1}$ and $v_{n-1}$ such that 
\begin{equation}
     \left( \prod\limits_{j=1}^{n-1} \ell_j \right) u_{n-1} - k_{n-1} v_{n-1} = \delta \, . 
\end{equation}
Using these integers, recall the shorthand notation $q_i = \ell_i v_{i-1}$. Finally, for $i=1,\dots,n-2$, let 
\begin{subequations}
\begin{equation}
    U_i := \left(\begin{array}{c|cc|c}
I_{i-1} &&\\ \hline
& u_{i}  & k_i \\
& v_i & \prod_{j=1}^{i}  \ell_{j}  \\ \hline 
& & & I_{n-i-1}
\end{array}\right) \in \mathrm{GL}(n,\mathbb{Z}) \, , 
\end{equation}
and 
\begin{equation}
    U_{n-1} := \left(\begin{array}{c|cc}
I_{n-2} &&\\ \hline
& u_{n-1}  &  \delta^{-1} k_{n-1}  \\
& v_{n-1} & \delta^{-1} \prod_{j=1}^{n-1}  \ell_{j}  
\end{array}\right)  \in \mathrm{GL}(n,\mathbb{Z})   \, ,  
\end{equation}
and 
\begin{equation}
    V = \left(\begin{array}{cccc|cc}
   1 & & & & & \\ 
   -q_2 & 1  & & & & \\ 
   +q_2 q_3 & - q_3 &1  & & & \\ 
- q_2 q_3 q_4  & \ddots& \ddots & \ddots & &  \\ \hline
& & & & 1  &  0  \\
& & & & 0 & 1
\end{array}\right) \in \mathrm{GL}(n,\mathbb{Z})  \, . 
\end{equation}
\end{subequations}
In all the matrices here and below, the entries that are left blank are $0$. 

This is exactly what is needed to make the charge matrix lower diagonal by right multiplication: 
\begin{equation}
    \rho U_1 U_2 \cdots U_{n-1} V = \left(\begin{array}{c|cc}
I_{n-2} &&\\ \hline
* & \delta  &  0  \\
* & \epsilon  & 0  
\end{array}\right) \label{eq:upperDiag}
\end{equation}
In other words, we have reduced the quiver by redefining the gauge groups, in such a way that one U(1) factor now acts trivially, as shown by the last column which vanishes. The stars in the lower left block are integers which are not necessarily vanishing, but the exact values, which can be computed, are not important for the rest of the reasoning. The only important value is the entry denoted $\epsilon$. An explicit computation of the product reveals that $\epsilon = -q_n + k_{n-1}^{-1} u_{n-1} \prod_{i \in \mathbb{Z}_n} k_i $. Using the relations \eqref{eq:length_def} and \eqref{Eq:Cycle_Minimal_Conditions}, one checks that this expression simplifies into $\epsilon = \mathrm{gcd} (\ell_n , k_{n-2})$. As a consequence, we also have $\mathrm{gcd}(\delta , \epsilon)=1$. 

We are now almost ready to identify the mirror. The method outlined in Section \ref{Sec:Mirror_Construction} shows that it is enough to find a matrix $W \in \mathrm{GL}(n,\mathbb{Z})$ such that $W \rho U_1 U_2 \cdots U_{n-1} V $ is diagonal. This is achieved by 
\begin{equation}
    W = \left(\begin{array}{c|cc}
I_{n-2} &&\\ \hline
* & \$  &  \$  \\
* & \epsilon  & \delta  
\end{array}\right)
\end{equation}
where the stars denote coefficients chosen to cancel the stars in \eqref{eq:upperDiag}, and the $\$ $ signs denote Bézout coefficients that ensure the determinant is 1. We then get 
\begin{equation}
    W \rho U_1 U_2 \cdots U_{n-1} V = \left(\begin{array}{c|cc}
I_{n-2} &&\\ \hline
 & 1  & 0  \\
 & 0 & 0
\end{array}\right) \, ,
\end{equation}
which is simply the Smith reduction of the initial charge matrix.

The ones on the diagonal indicate that the gauge group of the mirror has no discrete factor, so it is exactly U(1). The charges of the hypermultiplets under this U(1) are the entries in the last row of the matrix $W$. We have not worked out all of them, but we know two of them, namely $\delta = \gcd (\ell_1 , k_{n-1})$ and $\epsilon = \gcd (\ell_n , k_{n-2})$. But the whole computation made here is invariant under a global shift by $\mathbb{Z}_n$. This means that the $n$ charges are precisely the $\gcd (\ell_{i+2} , k_{i})$ for $i \in \mathbb{Z}_n$. This concludes the proof of the Lemma in Section \ref{Sec:Abelian_Cycle_Minimal}.

\bibliographystyle{JHEP}     
\setstretch{1}
 \bibliography{references}

\end{document}